\documentclass[journal,draftclsnofoot,onecolumn,12pt]{IEEEtran} 

\usepackage{cite}
\usepackage{graphicx}
\usepackage{psfrag}
\usepackage[cmex10]{amsmath}
\usepackage{amssymb}
\usepackage{url}
\usepackage{subfigure}
\usepackage{algorithmic}
\usepackage{algorithm}
\usepackage{array}
\newcommand{\bs}[1]{\boldsymbol{#1}}
\newcommand{\mt}[1]{\text{#1}}
\newcommand{\mc}[1]{\mathcal{#1}}  
\newcommand{\mb}[1]{\mathbf{#1}}   
\newcommand{\eq}[1]{\begin{equation}{#1}\end{equation}}

\begin{document}

\title{Dynamic Radio Resource Management for Random Network Coding: Power Control and CSMA Backoff Control}

\author{\IEEEauthorblockN{Kai Su, Dan Zhang, Narayan B. Mandayam}
\IEEEauthorblockA{ \\WINLAB, Rutgers University\\
671 Route 1 South, North Brunswick, NJ 08902 \\
Email: \{kais, bacholic, narayan\}@winlab.rutgers.edu}
\thanks{*This work is supported in part by the NSF under grant no. CCF-1016551. This work was presented in part at CISS 2012 and ISIT 2012.}
}
%
\maketitle

\vspace{-1.3cm}
\begin{abstract}
Resource allocation in wireless networks typically occurs at PHY/MAC layers, while random network coding (RNC) is a network layer strategy. An interesting question is how resource allocation mechanisms can be tuned to improve RNC performance.
By means of a differential equation framework which models RNC throughput in terms of lower layer parameters, we propose a gradient based approach that can dynamically allocate MAC and PHY layer resources with the goal of maximizing the minimum network coding throughput among all the destination nodes in a RNC multicast. We exemplify this general approach with two resource allocation problems:  (i) power control to improve network coding throughput, and (ii) CSMA mean backoff delay control to improve network coding throughput. We design both centralized algorithms and online algorithms for power control and CSMA backoff control. Our evaluations, including numerically solving the differential equations in the centralized algorithm and an event-driven simulation for the online algorithm, show that such gradient based dynamic resource allocation yields significant throughput improvement of the destination nodes in RNC. Further, our numerical results reveal that network coding aware power control can regain the broadcast advantage of wireless transmissions to improve the throughput.
\end{abstract}

\begin{IEEEkeywords}
Resource allocation, differential equation framework, random network coding, power control, CSMA.
\end{IEEEkeywords}

\newpage

\section{Introduction}
In wireless networks, resource allocation could take place at multiple layers of the protocol stack. Examples of these include transmit power control, channel allocation, and link scheduling at the PHY/MAC layer and buffer management at the transport layer. While network layering aims to reduce inter-layer dependency and brings noticeable benefits for interconnection, it is recognized that network performance can be optimized if network resources at different layers can be jointly taken into consideration for the design of network information flow. Specifically, the PHY and MAC layer resources, which tend to be isolated from upper layer functionalities, can be designed to support performance requirements at routing and transport layers \cite{tan-pc}. The resource allocation problem has been extensively studied for different types of wireless networks (see \cite{geo-realloc,xia-realloc-suv,mung-realloc-suv}), such as cellular networks and wireless ad hoc networks. Nevertheless, the methodologies adopted in these works, for instance, maximum network utility, were built upon the premise that networks leverage traditional routing and forwarding. Therefore, when a different transport paradigm, such as random network coding (RNC), is employed, the traditional approaches fail to take full advantage of the distinct properties of RNC, and thus become less effective or even invalid. 

RNC allows the nodes in the network to perform coding of packets at the network layer. It has received a large amount of attention since its inception \cite{netflow} and  has been demonstrated to yield benefits in achieving the optimal network throughput \cite{netflow}, improving network security \cite{han-sec}, and supporting distributed storage \cite{dima-dist-stor} and content delivery \cite{gkan-cont-dist}. Compared with the active progress in researching the benefits and application of RNC, there have been far less efforts looking into resource allocation for RNC. Existing ones include \cite{yal-cros,dan-rand,zha-reallocisit}. In fact, due to the cooperative nature of RNC, varied allocation of resources at different nodes lead to complex interactions and unpredictable performance. We will elaborate on this complex interaction using two motivating examples: (i) power control in a wireless network with RNC, and (ii) CSMA backoff control in a wireless network with RNC. 

\begin{figure}[h]
\begin{center}
\includegraphics[height=4cm,width=11cm]{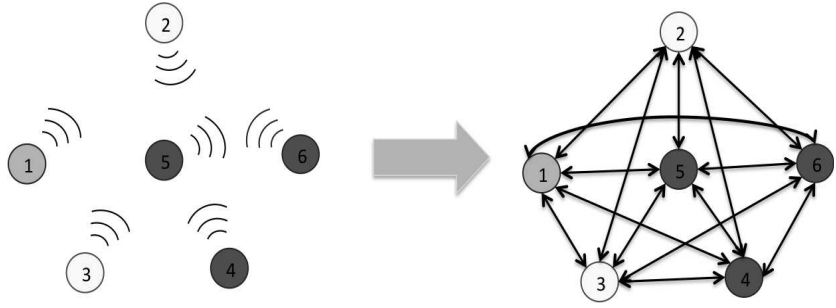}
\end{center}
\caption{Hypergraph model of a wireless network of six nodes with $s=1$ and $\mathcal{D}=\{4,5,6\}$.}
\label{fig:concept_topo}
\end{figure}

\begin{figure}[h]
\begin{center}
\includegraphics[height=8cm]{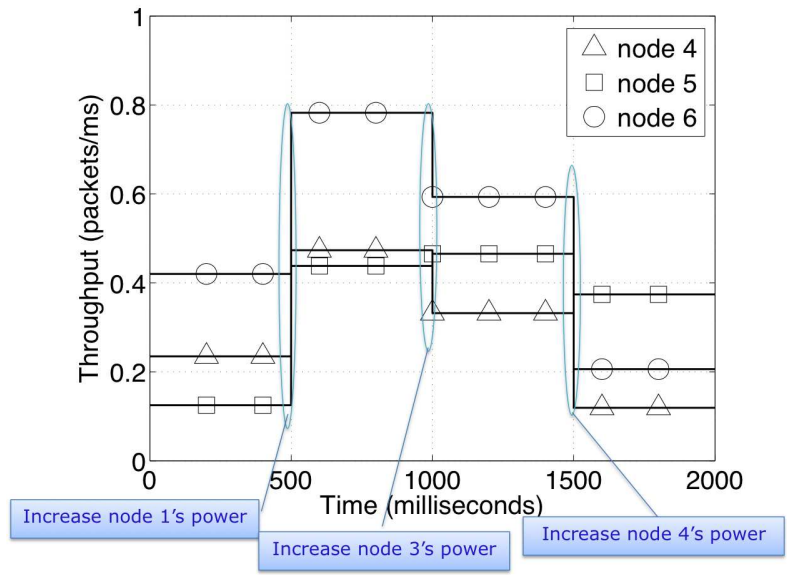}
\end{center}
\caption{Effect of transmit power on network coding throughput.}
\label{fig:moti-p}
\end{figure}

Let us first consider the effects of transmit powers on the performance of random network coding in the wireless network shown in Figure \ref{fig:concept_topo}. The source node, node $1$, is trying to multicast to a set of sink nodes, node $\{4,5,6\}$. In this network, every node is transmitting and is also able to receive from others. We further assume the network is interference limited, i.e., each transmission is interfered by simultaneous transmissions from other nodes. Therefore, increasing transmit power at a node improves SINR value of its own transmission but raises interference to others. The throughput of the destination nodes thus depend on the power levels at all nodes. To observe this effect, we set the transmit power $P_{i}^{\mt{Tx}}$ of each node $i$ to $13$dBm at $t=0\text{ms}$. Subsequently, at $t=500\text{ms}, t=1000\text{ms}$ and $t=1500\text{ms}$, the transmit powers of node $1, 3, 4$, i.e., $P_{1}^{\text{Tx}}, P_{3}^{\mt{Tx}}$ and $P_{4}^{\mt{Tx}}$ are increased to $14$dBm, respectively. As seen in Figure \ref{fig:moti-p}, the power increment of node $1$ at $500\mt{ms}$ improves the throughput of all the sink nodes, whereas node $3$'s increment at $1000\mt{ms}$  leads to the decrease of throughput of node $4$ and $6$. Therefore, increasing power at one node does not necessarily improve the throughput at all the destination nodes; on the contrary, it may possibly hurt the throughput at some node. 

\begin{figure}[h]
\begin{center}
\includegraphics[height=8cm]{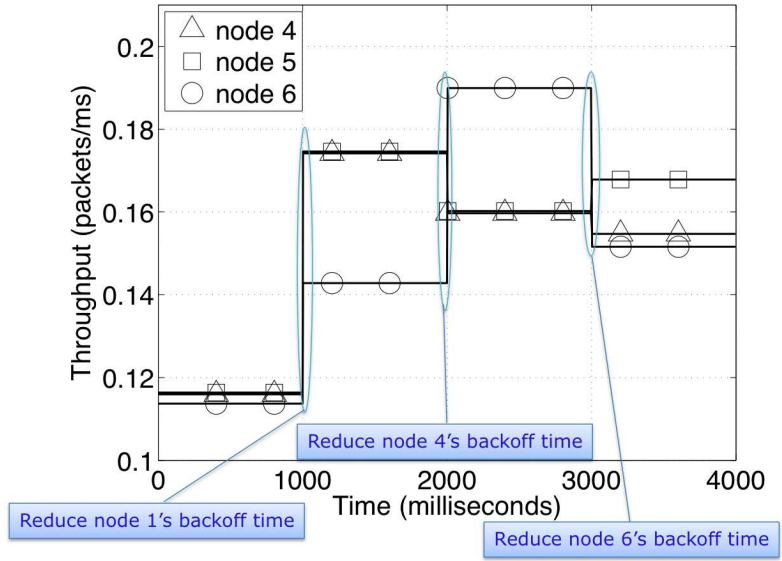}
\end{center}
\caption{Effect of CSMA backoff delay on network coding throughput.}
\label{fig:moti-c}
\end{figure}

Now consider the case of adjusting the backoff time in a CSMA network employing RNC. We consider a network with the same topology as in Figure \ref{fig:concept_topo} that is  utilizing CSMA as the MAC layer protocol. In this network, each node contends for transmission with an exponentially distributed delay value. We manipulate the mean of the backoff delay to control the transmission aggressiveness of each node and see its impacts on RNC throughput. 
At $t=0\mt{ms}$, the mean backoff delay of each node is set such that all the destination nodes, node $4, 5, 6$, are transmitting at about $0.12\text{pkt/ms}$. Subsequently, at $t=1000\text{ms}, t=2000\text{ms}$ and $t=3000\text{ms}$, the mean backoff delay of nodes $1, 4, 6$ are reduced, i.e., transmission aggressiveness increased, respectively as follows. The mean backoff delay of node $1$ is reduced from $3.70$ms to $2.24$ms, node $4$ from $2.74$ms to $1.66$ms, and node $6$ from $1.66$ms to $0.83$ms. Figure \ref{fig:moti-c} shows that, for example, at $t=2000\mt{ms}$, when node $4$ starts to contend more aggressively, it improves the throughput of node $6$. However, this simultaneously leads to the drop of the throughput of node $4$ itself and node $5$. An apparent reason is that it leads to reduced channel availability for these two nodes. Similar effects can also be seen for the subsequent window size change when node $6$ becomes more aggressive. 


Both of the above two examples, one at the PHY layer, and the other at the MAC layer, show that due to the network dynamics and the complexity of the problem, it would be cumbersome or unsuccessful to employ some static, or heuristic resource allocation mechanism in a network employing RNC. Rather, a deliberately designed, and more importantly, dynamic resource allocation algorithm is required to support the optimal RNC performance. 
Since RNC is fundamentally different from routing and forwarding in terms of packet delivery  as there are no specific routes being computed and followed \cite{bene}, analyzing it with traditional methods designed for uncoded networks will be problematic. 
To illustrate this, note that in a network where information is delivered based on some proactively determined route, network resources should be arranged such that certain network throughput can be achieved with certain reliability. However because of the nature of RNC, reliability is not a separate issue to be taken care of; resource allocation to improve network coding performance mainly is about improving throughput. Therefore, adopting an inappropriate model will not take full advantage of the benefits that RNC offers. In light of this, a differential equation based framework in \cite{dedi-it} and \cite{dedi-isit} is of particular interest for deriving resource allocation algorithms for RNC. This framework leverages a system of differential equations to elegantly model the rank evolution process which shapes the RNC performance. The presence of PHY and MAC layer parameters in this model makes it natural to analyze lower layer resource allocation for RNC.


In this paper, we address the problem of resource allocation for random network coding in wireless networks.  In what follows, we first discuss the system model considered in this article and analyze RNC throughput with the differential equation framework in section \ref{sec:sys-mod}. Then in section \ref{sec:algo}, we formulate the resource allocation problem to maximize the minimum network throughput among the destination nodes. While this problem falls into the category of optimal control, which is usually solved by the method of calculus of variations, we present a gradient based algorithm specifically designed for this resource allocation problem here. Two use cases of this algorithm, i.e., power control and CSMA mean backoff delay control, are presented to improve network coding throughput in section \ref{sec:pc} and \ref{sec:csma}, respectively. Specifically, we derive both centralized and online algorithms for the above two use cases. The main contribution of this work is to present a novel methodology to analyze cross-layer resource allocation in the context of RNC from a dynamical system view provided by the differential equation model. The framework utilized in this methodology is sufficiently general such that it can be used to analyze all kinds of PHY/MAC layer resources and derive effective resource allocation algorithms.


\section{Preliminaries} \label{sec:sys-mod}

\subsection{Differential Equation Framework for RNC} \label{inter-mod}
We adopt the directed hypergraph, introduced in \cite{oncoding}, to model a wireless network: $G=(\mathcal{N}, \mathcal{E})$ which has $N$ nodes $ \mathcal N= \{ 1,2, \ldots , N \} $ and hyperarcs $\mathcal{E}=\{ (i,\mathcal{K} )|i \in \mathcal{N}, \mathcal{K} \subset \mathcal{N}\}$. By introducing the hyperarc $(i,\mathcal{K})$, this model naturally captures the fact that in a wireless environment, a packet transmitted by node $i$ can be received by a subset of nodes from $\mathcal{K}$. 
To illustrate this, we note from Figure \ref{fig:concept_topo} that each node has a point to point link to every other node, but a transmitted packet can only be received by a subset of these nodes. This subset could be determined explicitly by the received signal and interference levels (adopted in section \ref{sec:pc} for the case of power control), or implicitly by a thresholding distance for reception (adopted in section \ref{sec:csma} for the case of CSMA). 

\begin{figure}[t]
\label{fig:re1}
\begin{center}

	\subfigure[4-node network topology] {
		\label{fig:re1-t}
		 \includegraphics[height=4.5cm]{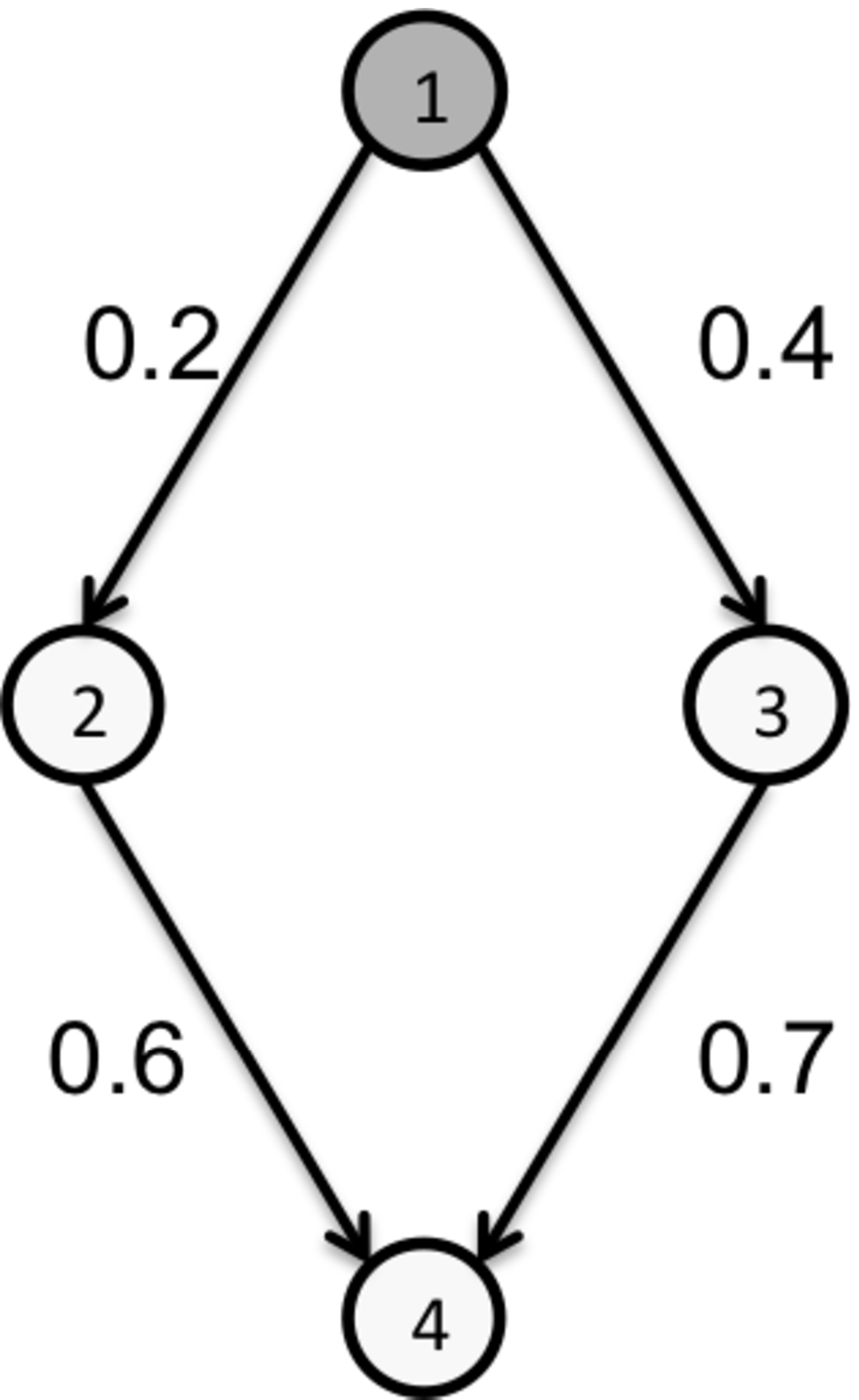}
	}
	\subfigure[Rank evolution] {
		\label{fig:re1-r}
		 \includegraphics[width=7.5cm]{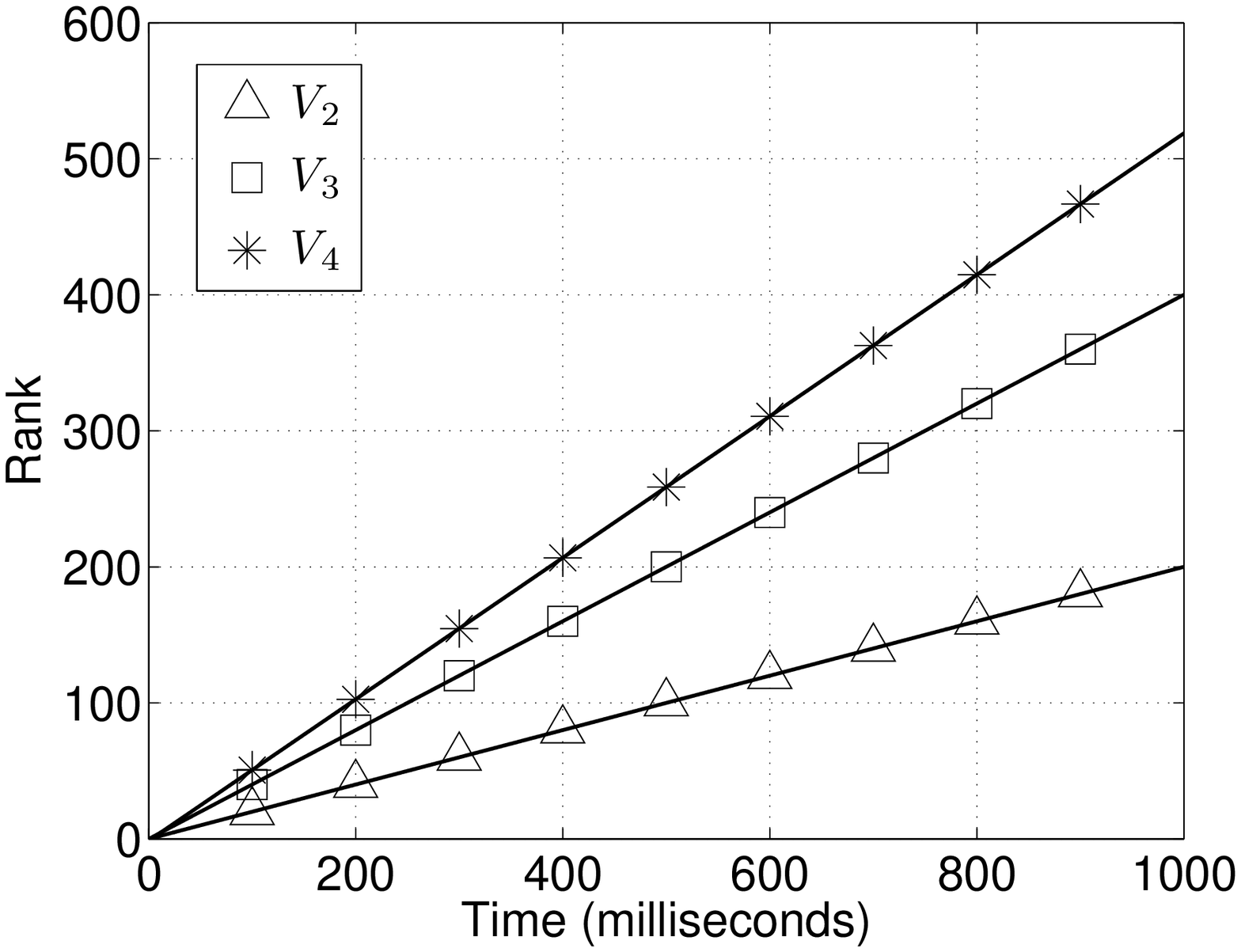}
	}
\caption{Rank evolution modeled by DE, example 1.}
\end{center}
 \vspace{-0.3cm}
\end{figure}

\begin{figure}[t]
\label{fig:re2}
\begin{center}

	\subfigure[8-node network topology] {
		\label{fig:re2-t}
		 \includegraphics[height=4.5cm]{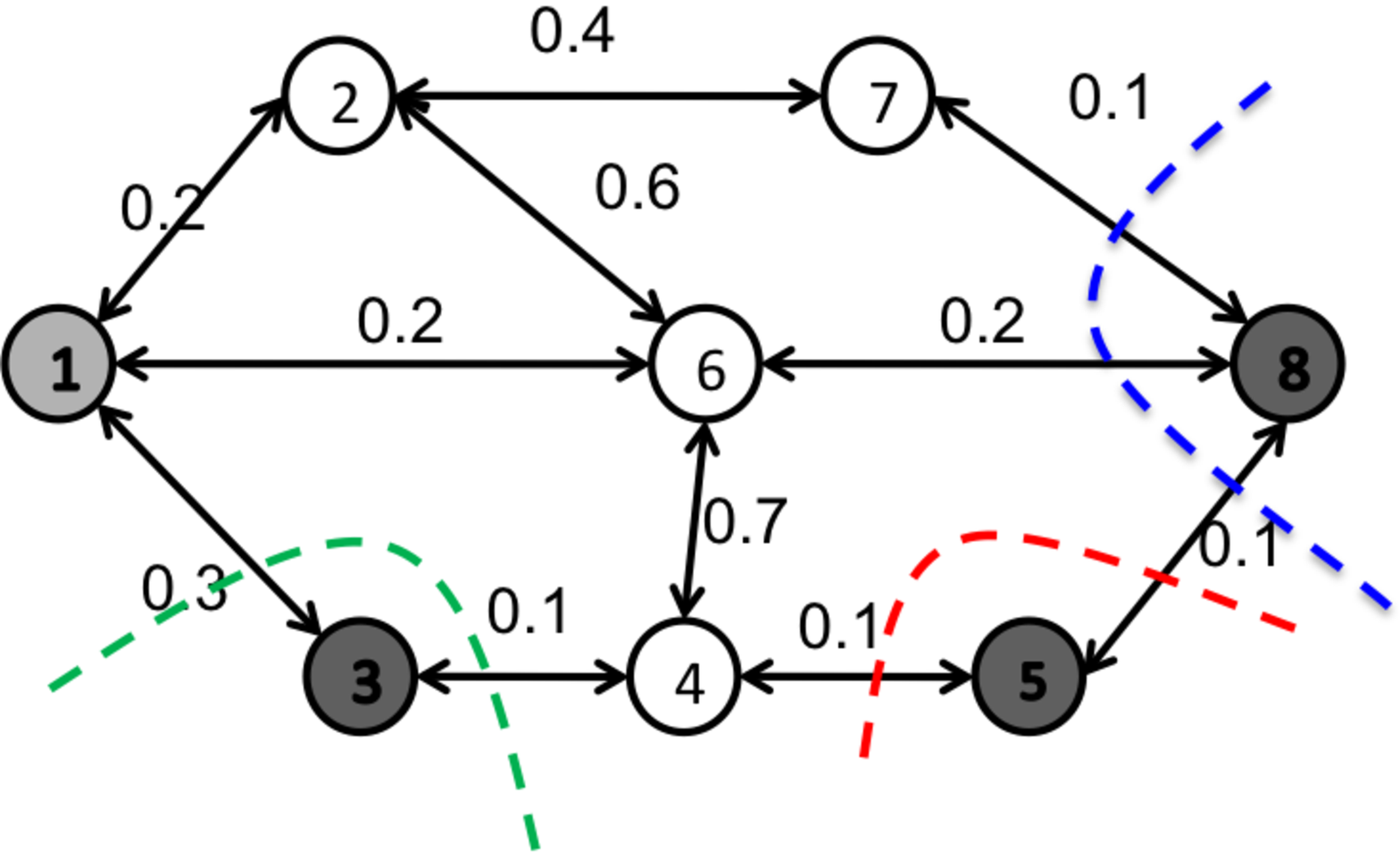}
	}
	\subfigure[Rank evolution] {
		\label{fig:re2-r}
		 \includegraphics[width=7.5cm]{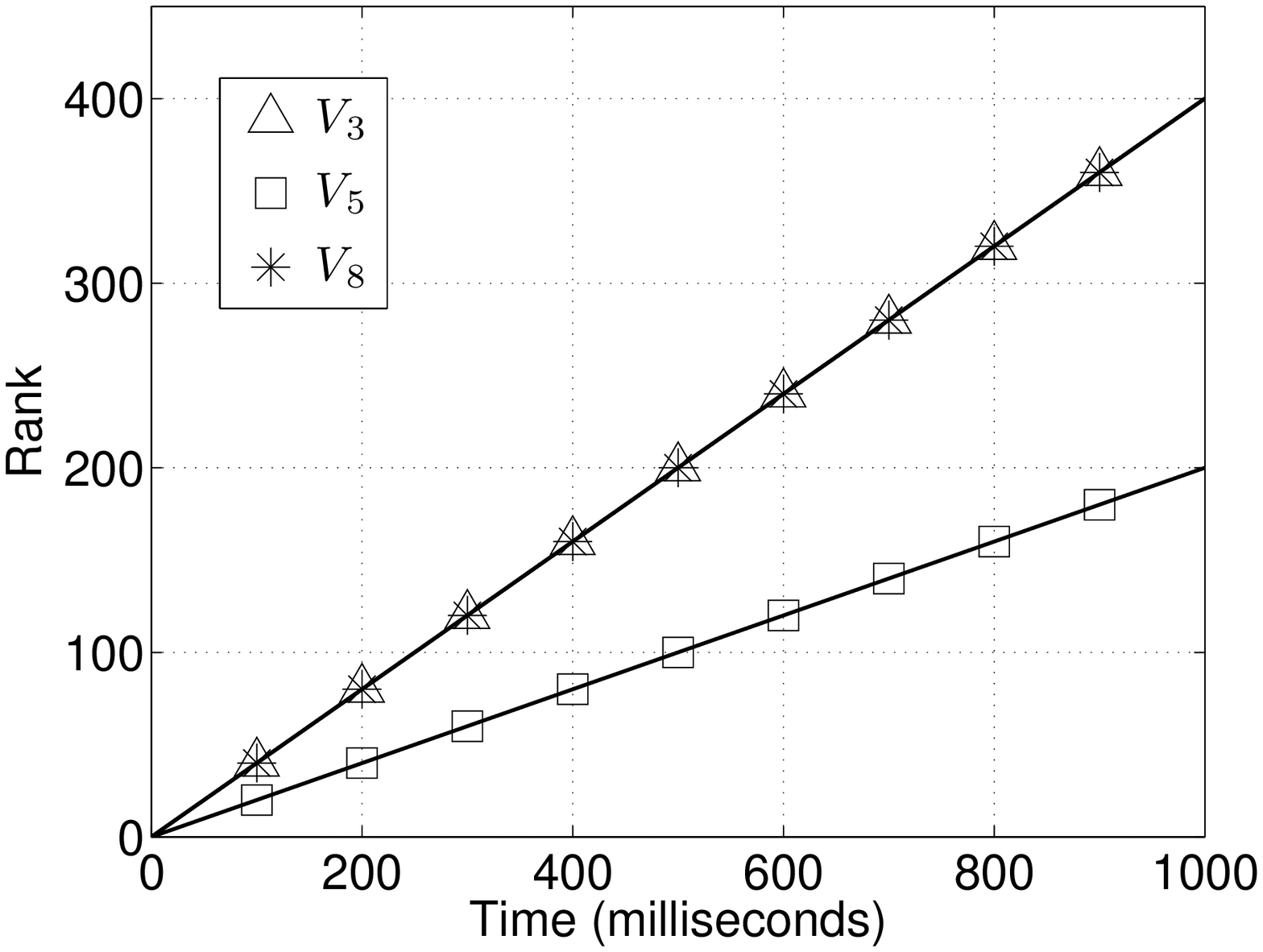}
	}
\caption{Rank evolution modeled by DE, example 2.}
\end{center}
 \vspace{-0.3cm}
\end{figure}

Consider that each node in the wireless network $G$ is performing random network coding \cite{bene}, i.e., a source node sends out random linear combinations of the original packets (coded packets), and other nodes merely receive packets from the network and they in turn send out random linear combinations of the packets received. We assume each coded packet is a row vector of length $l$ from $\mathbb{F}_q^l$, where $q$ is the field size. No routing operations are performed in the network and destination nodes can recover the original packets after collecting sufficient coded packets.  Assuming packet loss is only due to bit errors, we can then define the probability that a packet transmitted by node $i$ can be received by at least one node in $\mc{K}$, $P_{i,\mc{K}}$ as 
\begin{equation} \label{eq:pik} 	
P_{i,\mathcal{K}} = 1- \prod_{j \in \mathcal{K}} \left(1-P_{i,j}\right),
\end{equation}
where $P_{i,j}$ is the reception probability of link $(i,j)$. We can see that $P_{i,\mc{K}}$ is a function of the PHY layer parameters, e.g., transmit powers and interference. Assuming there exists certain MAC protocol running in its stable state such that  node $i$ is sending out coded packets at the average rate of $\lambda_i$ packets per second, then the successful transmission rate from node $i$ to node $j$ can be defined as
\begin{equation} \label{eq:zik}
z_{i,\mathcal{K}} = \lambda_i P_{i,\mathcal{K}}.
\end{equation}
$z_{i,\mathcal{K}}$ can also be regarded as the capacity of hyperarc $(i,\mc{K})$. Note that the capacity of a cut, $\mc T$ for $(\mc S, \mc K), \ \mc S, \mc K \subset \mc N$ where $\mc K \subset \mc T \subset \mc S^c$ is given as $c(\mathcal{T})=\sum_{i\in \mathcal{T}^c} z_{i,\mathcal{T}}$. Then, the min cut for $(\mathcal{S}, \mathcal{K})$ is the cut with the minimum size. 
We call the number of linearly independent coded packets the rank, and use $V_{\{i\}}$ to denote the rank at node $i$. We define an \textit{innovative} packet of node $i$ as the received packet which increases the rank $V_{\{i\}}$. For a RNC multicast session, if there are $m$ original packets to be delivered, each destination node $i$ can decode only if $V_{\{i\}} = m$. The notion of rank can be naturally extended to a set of nodes, $\mc{K}$, and thus $V_{\mc K}$ is the joint rank of all the nodes from $\mc K$. We call the stochastic process $V_{\mathcal{K}}(t)$ that grows from $0$ to $m$ the rank evolution process. In \cite{dedi-it}, it has been established that under the fluid approximation, we have a concentration result for the rank evolution process, i.e., the stochastic process $V_{\mathcal{K}}(t)$ is well represented by its mean, $E[V_{\mathcal{K}}(t)]$. Also, the following system of differential equations have been derived to model the rank evolution process $V_{\mc K}(t)$:
\begin{equation} \label{eq:dotv}
\dot{V}_{\mathcal{K}} = \sum_{i\notin \mathcal{K}} z_{i,\mathcal{K}}(1-q^{V_{\mc{K}}-V_{\{i\}\cup \mc{K}}}), \  \ \  \forall \mathcal{K} \subset \mathcal{N} \ \text{and}\ \mc{K}\ne \emptyset.
\end{equation}
It is worth noting that $\dot{V}_{\mathcal{K}}$ is the rate at which $\mc{K}$ is receiving innovative packets, i.e., $\dot{V}_{\mathcal{K}}$ denotes the throughput of $\mc{K}$. Apparently, with $z_{i,\mathcal{K}}$ being an abstraction of the outcome of all the PHY/MAC operations in the system of differential equations (\ref{eq:dotv}), the throughput of a set of nodes can be elegantly analyzed with respect to PHY/MAC parameters.  To illustrate this, we present two numerical examples. First consider a wireless network shown in Figure \ref{fig:re1-t} (also discussed in \cite{dedi-it}) where the source node, node $1$ intends to multicast $1000$ packets to destination nodes $2, 3, 4$. Let each node perform RNC operations and transmit at $1$pkt/ms. Suppose that packets from node $1$ can only be successfully received by node $2$ and $3$, with probability of $0.2$ and $0.4$, respectively, and node $2$ and node $3$'s packets can only be successfully received by node $4$ with probability of $0.6$ and $0.7$, respectively. Based on the above parameters, we can compute the successful transmission rate $z_{i,\mc K}$ for each hyperarc $(i, \mc K)$ for this topology. Then all the $z_{i,\mc K}$ are plugged in the system of differential equations (\ref{eq:dotv}) and solving them yields the result shown in Figure \ref{fig:re1-r}, the plot of rank evolution process for this RNC multicast. It can be easily verified that the throughputs of the destinations, i.e., the rates at which ranks increase, match the min cuts of every source and destination pair. For instance, it is trivial to see the min cut between node $1$ and $2$ is $0.2$, which is equal to the slope of the straight line for $V_2$ in Figure \ref{fig:re1-r}. Next, we  present an example with a larger topology. Consider the eight-node wireless network shown in Figure \ref{fig:re2-t}. Each line connecting two nodes denotes a bidirectional communication link with the packets reception probability next to the line. This time node $1$ has $1000$ packets to deliver to node $3, 5$ and $8$. We still let each node transmit at $1$pkt/ms. The result of solving equations (\ref{eq:dotv}) for rank evolution is shown in Figure \ref{fig:re2-r}. Again, the system of DEs serve as an accurate analytical model of RNC throughput. The throughputs implied by Figure \ref{fig:re2-r}, i.e., throughputs of nodes $3, 5$ and $8$ being $0.4$pkt/ms, $0.2$pkt/ms, and $0.4$pkt/ms, respectively, match the values of min cuts highlighted by the dashed curves in Figure \ref{fig:re2-t}. Thus the DE framework in \cite{dedi-it} is versatile and can be used to study the dynamics of RNC in any arbitrary network. In this paper, we will develop a dynamic radio resource management methodology using this framework.

%

\section{Resource Allocation Algorithm \\for Wireless Network Coding} \label{sec:algo}

\subsection{Problem Formulation}
Consider a wireless network $G = (\mc N, \mc E)$ which is performing random network coding. The source node $s$ tries to multicast $m$ packets to a set of sink nodes. We proceed to consider some PHY or MAC layer resource at every node $i$ and denote it as $r_i$. Note that $r_i$ can be any PHY/MAC layer parameter which contributes to the transmission rate $\lambda_i$ or the packet reception probability $P_{i,\mc K}$. Letting the vector $\mb r$ denote $[r_1, r_2,..., r_N]^{\top}$, we have
\eq{ 
z_{i,\mc K}  = z_{i,\mc K}(\mb{r}),
}
i.e., the reception rate $z_{i,\mc K}$ for each hyperarc $(i,\mc K)$ is an explicit function of allocated resource $\mb r$. To formulate the resource allocation as an optimization problem for improving the RNC performance, we consider maximizing the minimum throughput among all the sink nodes as the objective function. We let $\mc R$ be the set of destination nodes which have not reached full rank, $m$. With a little abuse of notation, we let $\dot V_{\{i\}}$ denote $\dot V_i$. Then letting $k = \arg \min_{j\in \mathcal{R}} \dot{V}_j$, we construct an optimization problem to maximize $\dot V_k$ as follows: 
\eq{
 \label{eq:max-min}
\begin{aligned}
& \underset{}{\text{maximize}}
&& \dot V_k \\
& \text{subject to}
& & \dot{V}_{\mathcal{K}} = \sum_{i\notin \mathcal{K}} z_{i,\mathcal{K}} \cdot  (1-q^{V_{\mc{K}}-V_{\{i\}\cup \mc{K}}}), \ \forall \mathcal{K} \subset \mathcal{N}.\\
& & & z_{i,\mathcal{K}} = z_{i,\mathcal{K}}(\mb r).\\
& & & k = \arg \min_{j\in \mathcal{R}} \dot{V}_j \\
& \text{variables}
& & \mb r.
\end{aligned}
}

\subsection{Gradient-based Resource Allocation Algorithm}
Note that in general, the optimization problem given by (\ref{eq:max-min}) is not convex, and thus it is difficult to find the optimal solution efficiently. Additionally, this type of problem which is constrained by a set of first-order differential equations can be categorized into an optimal control problem. Existing approaches to optimal control involve calculus of variations, which can be computationally expensive and intractable in wireless networks. In this paper, we aim to find a local optimum of this problem which can provide significant throughput gains with less computational complexity. We take an approach based on the steepest ascent (its counterpart for minimization problems is called steepest descent, see \cite{rus-opt}), i.e. adjust the resource $\mb r$ towards the direction of the gradient. Essentially, our optimization objective, $\dot V_k$ is a function of the resource $\mb r$, i.e., $\dot V_k = \dot V_k (\mb r)$. This allows us to establish the gradient of throughput as the direction of the dynamic adjustment of the resource, i.e.,
\eq{ \label{eq:rdot}
\dot{\mb r}= a^{\prime} \cdot \nabla \dot V_k,
}
where $a^{\prime}$ is a positive constant tuning the gain. In this way, the allocation of resource will be iterative, as well as dynamic. We consider a discrete approximation to compute the derivative in equation (\ref{eq:rdot}) as follows. Let $\Delta v$ be the step size, and $\mb e_i$ be a column vector with $1$ being the $i$th component and $0$ elsewhere. Writing in component-wise form, we have 
\eq{ 
\dot{r}_i = a^{\prime} \cdot \frac{\dot V_k(\mb r + \Delta v\mb e_i) - \dot V_k(\mb r)}
{\Delta v}.
}
Replacing $a^{\prime}/ \Delta v$ with $a$, we have:
\eq{ \label{eq:gen-algo}
\dot{r}_i = a \cdot \left( \dot V_k(\mb r + \Delta v\mb e_i) - \dot V_k(\mb r) \right).
}
Equation (\ref{eq:gen-algo}) serves as the basis of the resource allocation algorithm and works in an iterative manner to adapt the resource allocation towards the direction of the gradient of the minimum throughput. In the above, $\dot V_k$ is given by equation (\ref{eq:dotv}) and thus the allocation of resources takes into consideration the latest network throughput information and therefore also works in a dynamic fashion.

Note that until now, we have not imposed any specific models for the PHY/MAC layers. In fact, the resource allocation approach presented here is flexible enough that it can work with any specific lower layer models/mechanisms. For a better elucidation of this approach, we illustrate its applicability by solving two practical allocation problems for RNC: (i) power control for maximizing the minimum throughput, and (ii) CSMA mean backoff delay control for maximizing the minimum throughput. 
In what follows, we present the particular PHY/MAC layer models considered for the exemplary problems and derive the corresponding algorithms.

\section{Dynamic Power Control in RNC} \label{sec:pc}
There exists a rich history of transmit power control for cellular networks (see \cite{mun-pc}) where the goal was to minimize the total power levels \cite{fos-dpc,yates-pc,zand-pc}, or to maximize network utilities \cite{chiang-uti-pc,hand-utit-pc,mand-pc}. In this section, however, we consider performing power control in a coded wireless network to improve RNC throughput. 

\subsection{Interference Model} \label{inter-mod}
While the gradient based resource allocation framework in section \ref{sec:algo} is applicable for any wireless network with RNC, in this section we will specifically illustrate its use for power control in a network where we model the interference as Gaussian. In this section, we assume the wireless network $G$ to be interference limited, and model each point-to-point link gain $h_{ji}$ for $(i,j)$ with a path loss model. We use $P_{i}^{\mt{Tx}}$ to denote the transmit power at node $i$. The received signal level is given by $P_{i}^{\mt{Tx}}h_{ji}$. Each node $i$ is assumed to implement a certain processing gain $g_i$. Therefore, when node $j$ intends to receive the signal transmitted by $i$, the aggregated interference power is 
\begin{equation}
J_{ji} = \sum_{m\ne j,i}(P_{m}^{\mt{Tx}}\cdot h_{jm}/g_i).
\end{equation}
Let $\sigma^2$ denote the noise power. The signal-to-noise-and-interference ratio (SINR) for the point-to-point link $(i,j)$ can be written as
\begin{equation}
\text{\textit{SINR}}_{(i,j)} = \frac{P_{i}^{\mt{Tx}} \cdot h_{ji}}{J_{ji} +\sigma^2}.
\end{equation}

Assuming BPSK signaling and Gaussian interference, the bit error rate for sender-receiver pair $(i,j)$ is given as
\eq{ \label{ber}
p_{i,j}^{\text{bit}}=Q \left( \sqrt{ \mt{\textit{SINR}}_{(i,j)} }  \right).
}

Further, assuming each packet is of $l$ bits, the probability that node $j$ can receive a packet without error is 
\begin{equation}
P_{i,\{j\}} = (1-p_{i,j}^{\text{bit}})^l.
\end{equation}

Note that the differential equation framework requires the computation of $P_{i,\mc K}$ given in equation (\ref{eq:pik}). Under the above interference model, we assume that there is a MAC protocol running in steady state such that each node $i$ has an average transmission rate $\lambda_i$. Therefore, $z_{i,\mc K}$ in equation (\ref{eq:zik}) can be now written as:
\eq{
\begin{array}{ll}
z_{i,\mc K} & = \lambda_i P_{i, \mc K} \\
&=\lambda_i \cdot \left(1- \prod_{j \in \mathcal{K}} \left(1- \left(1-Q\left(\sqrt{\frac{P_{i}^{\mt{Tx}} \cdot h_{ji}}{\sum_{m\ne j,i}(P_{m}^{\mt{Tx}}\cdot h_{jm}/g_i) +\sigma^2}
}
\right)\right)^l \right)\right).
\end{array}
}
\subsection{Centralized Power Control} \label{subsec:pcc}
Let $\mb P^{\mt {Tx}} = \left[P_{1}^{\mt{Tx}},P_{2}^{\mt{Tx}},...,P_{N}^{\mt{Tx}}\right]^\top$ be the transmit power vector and the resource $r_i = P_{i}^{\mt{Tx}}$. The centralized power control algorithm follows directly from equation (\ref{eq:gen-algo})  (see also \cite{ciss} and \cite{zha-reallocisit}):
\eq{ \label{eq:cpc}
\dot P_{i}^{\mt{Tx}} = a \cdot \left( \dot V_k(\mb P^{\mt{Tx}} + \Delta v\mb e_i) - \dot V_k(\mb P^{\mt{Tx}}) \right),
}
where
\eq{
\dot V_k = \sum_{i\ne k} 
\lambda_i \cdot  \left(1-Q\left(\sqrt{\frac{P_{i}^{\mt{Tx}} \cdot h_{ki}}{\sum_{m\ne k,i}(P_{m}^{\mt{Tx}}\cdot h_{km}/g_i) +\sigma^2}}
\right)\right)^l 
\cdot \left( 1-q^{V_k - V_{\{i,k\}}} \right),
}
based on the interference model above.

We assume there is a certain power budget at each node $i$, i.e., $0 \le P_{i}^{\mt{Tx}} \le P_{i}^{\mt{max}}$. Considering this, the algorithm can be summarized as:
\eq{ 
\boxed{
\left.
\begin{array}{l}
k = \arg \min_{j\in \mathcal{R}} \dot{V}_j\\
\dot P_{i}^{\mt{Tx}} = \left \{ \begin{array}{ll}
0,  
&  \text{if} \ (P_{i}^{\mt{Tx}}=P_{i}^{\mt{max}}\ \text{and}\ \dot V_k(\mb P^{\mt{Tx}} + \Delta v\mb e_i) > \dot V_k(\mb P^{\mt{Tx}}))\  \\ 
& \text{or }  (P_{i}^{\mt{Tx}}=0\ \text{and}\ \dot V_k(\mb P^{\mt{Tx}} + \Delta v\mb e_i) < \dot V_k(\mb P^{\mt{Tx}})); \\
a \cdot [\dot V_k(\mb P^{\mt{Tx}} + \Delta v\mb e_i) - \dot V_k(\mb P^{\mt{Tx}})], & \mt{otherwise.}
\end{array} \right. \end{array} \right.
}
}
To compute the throughput gradient for node $i$, the above algorithm requires the knowledge of $\dot V_k(\mb P^{\mt{Tx}} + \Delta v\mb e_i)$, the throughput of node $k$ when only node $i$'s power is incremented and other nodes' powers remain unchanged. This is only possible when there is a central controller which has the global knowledge of the whole network, especially the PHY layer specifics, such that $z_{i,k}$ can be obtained for all $i$, and further, the central controller needs to compute the power adjustment for each node. Therefore, we call it a \textit{centralized} resource allocation algorithm. This algorithm serves as a theoretical basis for the gradient algorithm considered here, but is difficult to be practically implemented to operate online, as it assumes the knowledge of PHY layer specifics all the time. 

\subsection{Online Power Control}
While the algorithm in section \ref{subsec:pcc} is effective in achieving the objective to improve network coding throughput \cite{zha-reallocisit}, it can be further improved for the purpose of implementation. Specifically, it would be more desirable if the algorithm can operate online, i.e., each node computes the gradient for itself and adjusts its power without any noticeable delay. 

We begin by discretizing the algorithm from \ref{subsec:pcc}, i.e., we divide the time into equal-length intervals, and seek an online algorithm which estimates the gradient at each interval, and adjusts powers at the end of each interval based on the most recent estimates. Let $T$ denote the objective function in equation (\ref{eq:max-min}), i.e., $T=\min_{j \in \mc R} \dot V_j$ and $\nabla_{\mathbf{P}^{\mt Tx}}T$ denote the gradient of $T$. The online algorithm is now given as:
\eq{
\mb{P}^{\mt {Tx}}(n) = \mb{P}^{\mt {Tx}}(n-1) + a \cdot \nabla_{\mathbf{P}^{\mt{Tx}}}T(n-1), \ \ n=2,3,4,...
}
where $a$ is the gain parameter for power control, and $\mb{P}^{\mt{Tx}}(n)$ and $T(n)$ are $\mb P^{\mt{Tx}}$ and $T$ in the $n$th interval, respectively. Note that the gist of the online algorithm lies in the estimation of $\nabla_{\mathbf{P}^{\mt{Tx}}}T(n-1)$ at each node based on the network information available to it, which is described in the following. 
We first introduce the following notations: 
\begin{itemize}
\item Let $\dot{\mb{V}}$ denote the size-$(2^N-1)$ throughput vector, consisting of $\dot{V}_{\mathcal{K}}, \forall \mc{K}\subset \mc{N}, \mc K \ne \emptyset$. More precisely, $\dot{\mb{V}}$ is given by
\eq{\dot{\mb{V}}=\left[ \dot V_{\{1\}} \ \dot V_{\{2\}} \ \dot V_{\{1,2\}} \ \dot V_{\{3\}}\ ... \ \dot V_{\{1,2,...,N\}}\right]
}
\item Let $\mb{z}$ denote the size-$\left(N\cdot(2^N-1)\right)$ packet reception rate vector, consisting of $z_{i,\mc{K}}, \forall i \in \mc{N}, \mc{K} \subset{N}, \mc K \ne \emptyset$. For each $i$, $z_{i, \mc K}, \forall \mc K \subset \mc N$ is arranged similarly with $\dot{\mb V}$. Then $\mb{z}$ is given by concatenating $z_{i, \mc K}, \forall \mc K \subset \mc N, \mc K \ne \emptyset$ from $i=1$ to $i=N$:
\eq{ \label{eq:z}
\mb{z} = \left[ z_{1,\{1\}} \ z_{1,\{2\}} \ z_{1,\{1,2\}}\ ...\ z_{1,\{1,2,...,N\}}\ ... \ z_{N,\{1,2,...,N\}}\right]
}
\item Let $z^\prime_{i,\mc{K}}$  denote the innovative packet reception rate of hyperarc $(i,\mc K)$.
\item Let $\mb{z}^\prime$ denote the size-$\left( N\cdot(2^N-1) \right)$ innovative packet reception rate vector, consisting of $z^\prime_{i,\mc{K}}, \forall (i, \mc K) \in \mc E$. By replacing every $z_{i,\mc K}$ of $\mb z$ with $z^\prime_{i,\mc K}$, we get $\mb z^\prime$.
\end{itemize}

Note that $T$ is a function of $\dot{\mb{V}}$, i.e., $T = T(\dot{\mathbf{V}})$. We remark that the online algorithm we are deriving can accommodate any optimization objective as long as $T$ remains a differentiable function of $\dot{\mb{V}}$ \cite{zha-reallocisit}. Here we use the minimum destination throughput as an example to derive the algorithm, i.e., $T=\dot V_k$, where $k = \arg \min_{j\in \mathcal{R}} \dot{V}_j $. Since we have $\dot {\mb V} = \dot {\mb V}(\mb z)$ and $\mb z = \mb z(\mb P^{\mt{Tx}})$, then based on the chain rule, gradient of $T$ with respect to transmit powers $\mb{P}^{\mt{Tx}}$ can be derived as:
\begin{equation} \label{grad-orig}
\nabla_{\mathbf{P}^{\mt{Tx}}}T =  
\nabla_{\dot{\mathbf{V}}}T 
J_{\mathbf{z}}\dot{\mathbf{V}} 
J_{\mathbf{P}^{\mt{Tx}}}\mathbf{z},
\end{equation}
where $\nabla_{\dot{\mathbf{V}}}T$, $J_{\mathbf{z}}\dot{\mathbf{V}}$, and $J_{\mathbf{P}_{\mt T}}\mathbf{z}$ are the gradient of $T$ with respect to $\dot{\mb{V}}$, the Jacobian of $\dot{\mb{V}}$ with respect to $\mb{z}$, and the Jacobian of $\mb{z}$ with respect to $\mb{P}^{\mt{Tx}}$, respectively. Equation (\ref{grad-orig}) provides the exact formula for computing the gradient of $T$. To derive the online algorithm, we shall first investigate the first two terms together and then the last term on the right hand side. 
Since we have $T=\dot V_k$, we can get
\eq{
\nabla_{\dot{\mathbf{V}}}T 
J_{\mathbf{z}}\dot{\mathbf{V}} =
\nabla_{\mb z} T =
\nabla_{\mb z} \dot V_k.
}
 Referring to equation (\ref{eq:z}), and noting $\dot{V}_{k} = \sum_{i\notin k} z_{i,k} \cdot  (1-q^{V_{k}-V_{\{i,k\}}})$, we have
 \eq{
 \nabla_{\mb z} \dot V_k = \left[ 
 0, 1-q^{V_{k}-V_{\{1,k\}}}, 1-q^{V_{k}-V_{\{2,k\}}}, 0, ... , 0, 1-q^{V_{k}-V_{\{k-1,k\}}}, 0, ..., 0, 
 1-q^{V_{k}-V_{\{k+1,k\}}}, ...
 \right].
 }
Noting $\nabla_{\mb z} \dot V_k$ has $(N-1)$ non-zero components: $1-q^{V_{k}-V_{\{i,k\}}}, \forall i \in \mc N, i \ne k$, we would like to avoid evaluating $\nabla_{\mb z} \dot V_k$ since it requires the knowledge of $V_{\{i,k\}}$, the joint rank of node $i$ and $k$. Instead, note that $z_{i,\mathcal{K}}$ is defined as the packet reception rate, and $(1-q^{V_{\mc{K}}-V_{\{i\}\cup \mc{K}}})$ is the probability that the received packet is an innovative packet. Then the innovative packet reception rate $z^{\prime}_{i, \mathcal{K}}$ is given by $z^{\prime}_{i, \mathcal{K}} = z_{i,\mathcal{K}}\cdot  (1-q^{V_{\mc{K}}-V_{\{i\}\cup \mc{K}}})$. Therefore, we take an alternative approach to consider the derivative with respect to the innovative packet reception rate $\mb{z}^\prime$ instead of $\mb{z}$. With $\mb{z}^\prime$, we can rewrite equation (\ref{eq:dotv}) in a more compact form:
\begin{equation} 
\dot{V}_{\mathcal{K}} = \sum_{i\notin \mathcal{K}} z_{i,\mathcal{K}}^\prime,  \ \forall \mathcal{K} \subset \mathcal{N},
\end{equation}
and $\nabla_{\mb z ^\prime} \dot V_k$ can be derived accordingly as:
\eq{
\nabla_{\mb z ^\prime} \dot V_k =
\nabla_{\dot{\mathbf{V}}}T \cdot
J_{\mathbf{z}^{\prime}}\dot{\mathbf{V}} =
\left[
0,\ 1,\ 1,\ 0, \ ...,\  0, \ 1, \ 0,\  ..., \ 0, \ 1, \ 0, \ ... \ 0
\right].
}

Then the gradient of $T$ can be calculated by
\begin{equation} \label{grad}
\nabla_{\mathbf{P}^{\mt {Tx}}}T = 
\nabla_{\dot{\mathbf{V}}}T 
J_{\mathbf{z}^{\prime}}\dot{\mathbf{V}} 
J_{\mathbf{P}^{\mt{Tx}}}\mathbf{z}^{\prime}.
\end{equation}

Thus the evaluation of equation (\ref{grad}) boils down to the evaluation of $J_{\mathbf{P}^{\text{Tx}}}\mathbf{z}^\prime$. Note we would avoid the approach of directly evaluating it as this would again require that the underlying PHY layer specifics of all the nodes be known universally. Furthermore, computing the Jacobian, i.e., $J_{\mathbf{P}^{\text{Tx}}}\mathbf{z}^\prime$ can be computationally prohibitive. Rather, we numerically estimate this Jacobian at each iteration by following a similar approach that is adopted in Broyden's method \cite{broyden}. For that purpose, we need to discretize the current algorithm which uses a continuous time model for network coding. We thus consider the time divided into equal-length time intervals (corresponding to the time between successive power updates) and assume that each interval is of $\tau$ seconds. Then according to \cite{broyden}, we have 
\eq{ \label{broyden}
J_{\mathbf{P}^{\text{Tx}}}\mathbf{z}^{\prime} (n) = J_{\mathbf{P}^{\text{Tx}}}\mathbf{z}^{\prime}(n-1) + 
\frac{\Delta\mathbf{z}^\prime(n) - J_{\mathbf{P}^{\text{Tx}}}\mathbf{z}^{\prime}(n-1) \Delta \mathbf{P}^{\text{Tx}}(n)}{\| \Delta \mathbf{P}^{\text{Tx}}(n)\|^2}
\Delta \mathbf{P}^{\text{Tx}\top}(n),
}
where $\Delta\mathbf{z}^\prime(n)=\mathbf{z}^\prime(n) - \mathbf{z}^\prime(n-1)$ and $\Delta \mathbf{P}^{\text{Tx}}(n) = \mathbf{P}^{\text{Tx}}(n) - \mathbf{P}^{\text{Tx}}(n-1)$.

We can estimate the average innovative packets reception rate in each time interval which is of length $\tau$, i.e.,
\eq{
z^\prime_{i,j}(n) = c_{i,j}(n)/\tau,
}
where $c_{i,j}(n)$ is the number of innovative packets sent by node $i$ that node $j$ has received in the $n$th time interval.
Note that the granularity of power updates are based on this time interval. Now with the time being discretized, we can rewrite equation (\ref{grad}) to give the gradient in the $n$th interval:
\begin{equation} \label{grad-disc}
\nabla_{\mathbf{P}^{\mt{Tx}}}T(n) = 
\nabla_{\dot{\mathbf{V}}}T(n) 
J_{\mathbf{z}^{\prime}}\dot{\mathbf{V}}(n) 
J_{\mathbf{P}^{\mt{Tx}}}\mathbf{z}^{\prime}(n).
\end{equation}

Note to initiate the algorithm, we need to provide an initial estimate of $J_{\mathbf{P}^{\mt{Tx}}}\mathbf{z}^{\prime}$, i.e., $J_{\mathbf{P}^{\mt{Tx}}}\mathbf{z}^{\prime} (0)$ and randomly generate the first power update. To detect the convergence of the online algorithm, a threshold value, which can be an achievable and satisfactory throughput value, should be set. In any interval $i$, the numerical value of $T(i)$ is monitored and if it grows larger than the threshold, the algorithm will maintain the previous allocation; otherwise, the gradient will be estimated and power at each node will be adjusted. Moreover, to refrain from unduly large gradient update which might lead the algorithm to an unstable state, we observe the results given by (\ref{grad-disc}) and if the gradient is larger than a threshold $\mb{s}$, we normalize the gradient, i.e., 
\begin{equation}
\nabla_{\mathbf{P}^{\mt{Tx}}}T(n) = \left \{ \begin{array}{ll}
				\frac{\nabla_{\dot{\mathbf{V}}}T(n) 
J_{\mathbf{z}^{\prime}}\dot{\mathbf{V}}(n) 
J_{\mathbf{P}^{\mt{Tx}}}\mathbf{z}^{\prime}(n)}
				{\|\nabla_{\dot{\mathbf{V}}}T(n) 
J_{\mathbf{z}^{\prime}}\dot{\mathbf{V}}(n) 
J_{\mathbf{P}^{\mt{Tx}}}\mathbf{z}^{\prime}(n)\|}, & 
				 \hfill \text{if}\ \nabla_{\dot{\mathbf{V}}}T(n) 
J_{\mathbf{z}^{\prime}}\dot{\mathbf{V}}(n) 
J_{\mathbf{P}^{\mt{Tx}}}\mathbf{z}^{\prime}(n) \ge \mb{s}\\
				\nabla_{\dot{\mathbf{V}}}T(n) 
J_{\mathbf{z}^{\prime}}\dot{\mathbf{V}}(n) 
J_{\mathbf{P}^{\mt{Tx}}}\mathbf{z}^{\prime}(n), & \text{otherwise.} 
				\end{array} \right.	
\end{equation}


Finally, the transmit powers are adapted according to the following equation
\begin{equation} \label{cont}
\mb{P}^{\mt{Tx}}(n) = \mb{P}^{\mt{Tx}}(n-1) + a \cdot \nabla_{\mathbf{P}^{\mt{Tx}}}T(n-1),
\end{equation}
where $a$ is the gain parameter for power control.
We summarize this algorithm in Algorithm \ref{algo}.
\begin{algorithm} 
\caption{Online network coding aware power control algorithm}
\begin{algorithmic} \label{algo}
\STATE  $\tau\gets$ Power update granularity
\STATE $\tau_0\gets$ Time interval to make gradient computation
\STATE rand(N,1) $\gets$ length-$N$ random vector with each element larger than $-0.5$ and less than $0.5$
\STATE $\mb{c}(0) \gets \mb{0}$: counters for number of innovative packets.
\STATE $\mb{P}^{\mt{Tx}}(0)\gets\mb{P}^{\mt{init}}$
\STATE $J_{\mathbf{P}^{\text{Tx}}}\mathbf{z}^{\prime} (0) \gets {J_{\mathbf{P}^{\text{Tx}}}\mathbf{z}^{\prime}} ^{\mt{init}}$
\WHILE {$t \le \tau$}
	\STATE update $\mb{c}(0)$
\ENDWHILE
\STATE compute $\mb{z}^\prime(0)$
\STATE $\mb{P}^{\mt{Tx}}(1) = \mb{P}^{\mt{Tx}}(0) + \text{rand}(N,1)$ 
\STATE $n \gets 1$
\WHILE{$ n\tau\le t \le (n+1)\tau$}
	\STATE update $\mb{c}(n)$
	\IF{$t > (n+1)\tau-\tau_0$}
		\WHILE{$\nabla_{\mathbf{P}^{\mt{Tx}}}T(n)$ has not been computed}
		\STATE compute $\mb{z}^\prime(n)$
		\STATE $\Delta \mb{P}^{\mt{Tx}}(n) = \mb{P}^{\mt{Tx}}(n) - \mb{P}^{\mt{Tx}}(n-1)$
		\STATE $\Delta\mathbf{z}^\prime(n)=\mathbf{z}^\prime(n) - \mathbf{z}^\prime(n-1)$
		\STATE Compute $J_{\mathbf{P}^{\text{Tx}}}\mathbf{z}^{\prime} (n)$ as in (\ref{broyden})
		\STATE Compute gradient $\nabla_{\mathbf{P}^{\mt{Tx}}}T(n)$
		\STATE Set power $\mb P^{\mt{Tx}}(n+1) = \mb P^{\mt{Tx}}(n) + a \cdot \nabla_{\mathbf{P}^{\mt{Tx}}}T(n)$
		\STATE $n\gets n+1$
		\ENDWHILE
	\ENDIF
\ENDWHILE
\end{algorithmic}
\end{algorithm}

\subsection{Numerical Results}
We use two ways to evaluate the power control algorithm for random network coding: for the centralized algorithm, we use a numerical differential equation solver to compute the throughput; and for the online algorithm, we implement the algorithm with an event-driven simulation. 

For the evaluations, we consider the topology of a 6-node network as shown in Figure \ref{fig:concept_topo}. This network is assumed to be running random network coding, i.e., original packets are being coded at each node, and the intended destination nodes decode to recover them. Therefore, the original packets are not ``routed" towards the destination nodes. The goal is for a sender node, node 1, to multicast $2000$ packets to a set of sink nodes, node $4, 5, 6$. We assume there is a path loss model for each point to point link in this network. We use the ITU model for indoor propagation \cite{itu}, where the path loss of the transmission from node $i$ to node $j$, $\text{PL}_{ji}$ is given by:
\eq{
\text{PL}_{ji} = 20\log f +10n \log d_{ji} + P_f(n) - 28.
}
In this evaluation, the transmitted signal frequency $f$ is set to be $2.4$GHz, path loss exponent $n$ to be 3,  and floor penetration factor $P_f(n)$ to be 11. Thus the link gain for $(i,j)$ can be computed as $h_{ji} = 10^{\text{PL}_{ji}/10}$.

\subsubsection{Centralized algorithm}
\begin{figure}[t]
\label{fig:pcc}
\begin{center}
	\subfigure[Effect of power control on throughput] {
		\label{fig:pcc-t}
		 \includegraphics[width=7.5cm]{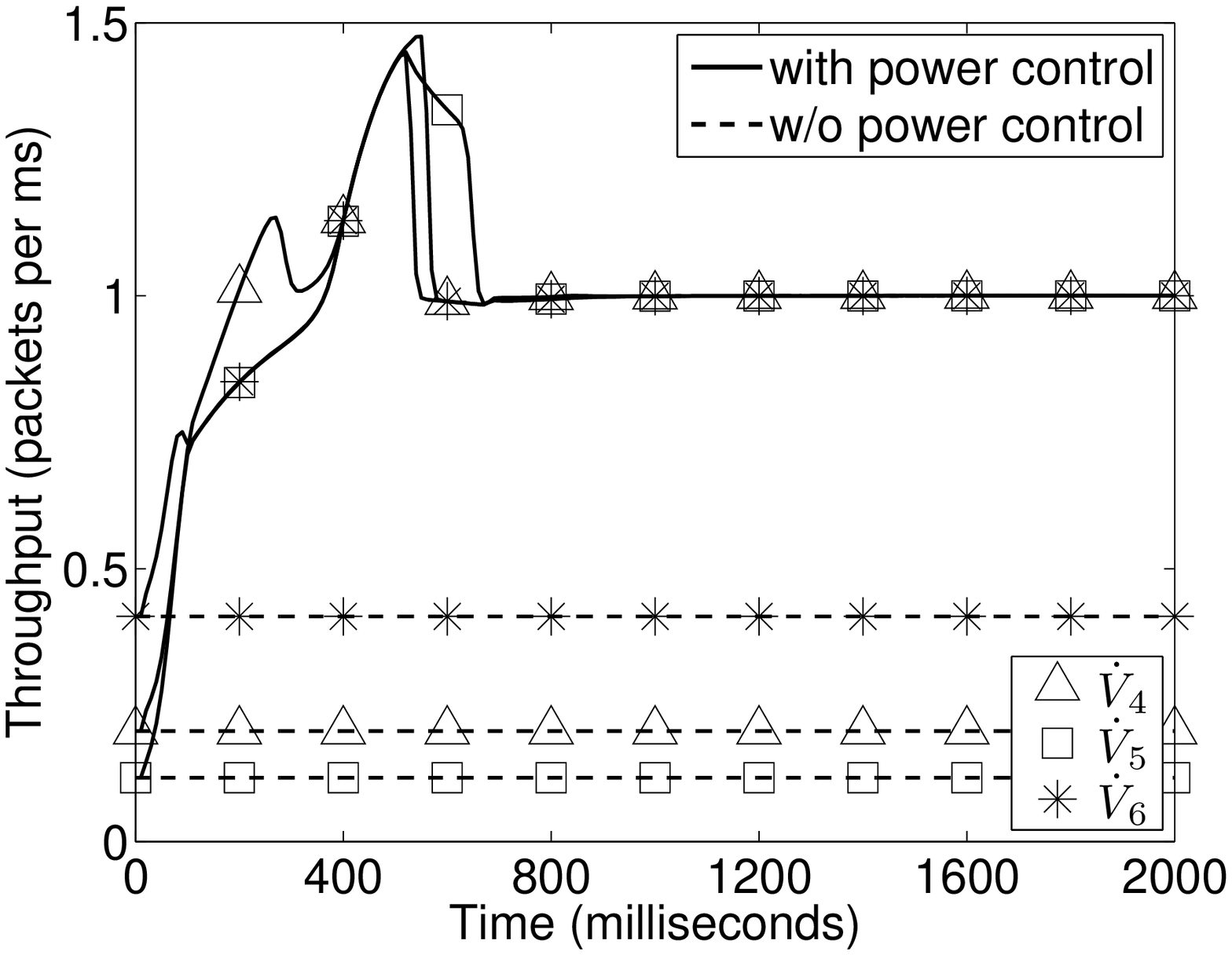}
	}
	\subfigure[Power adjustment] {
		\label{fig:pcc-p}
		 \includegraphics[width=7.5cm]{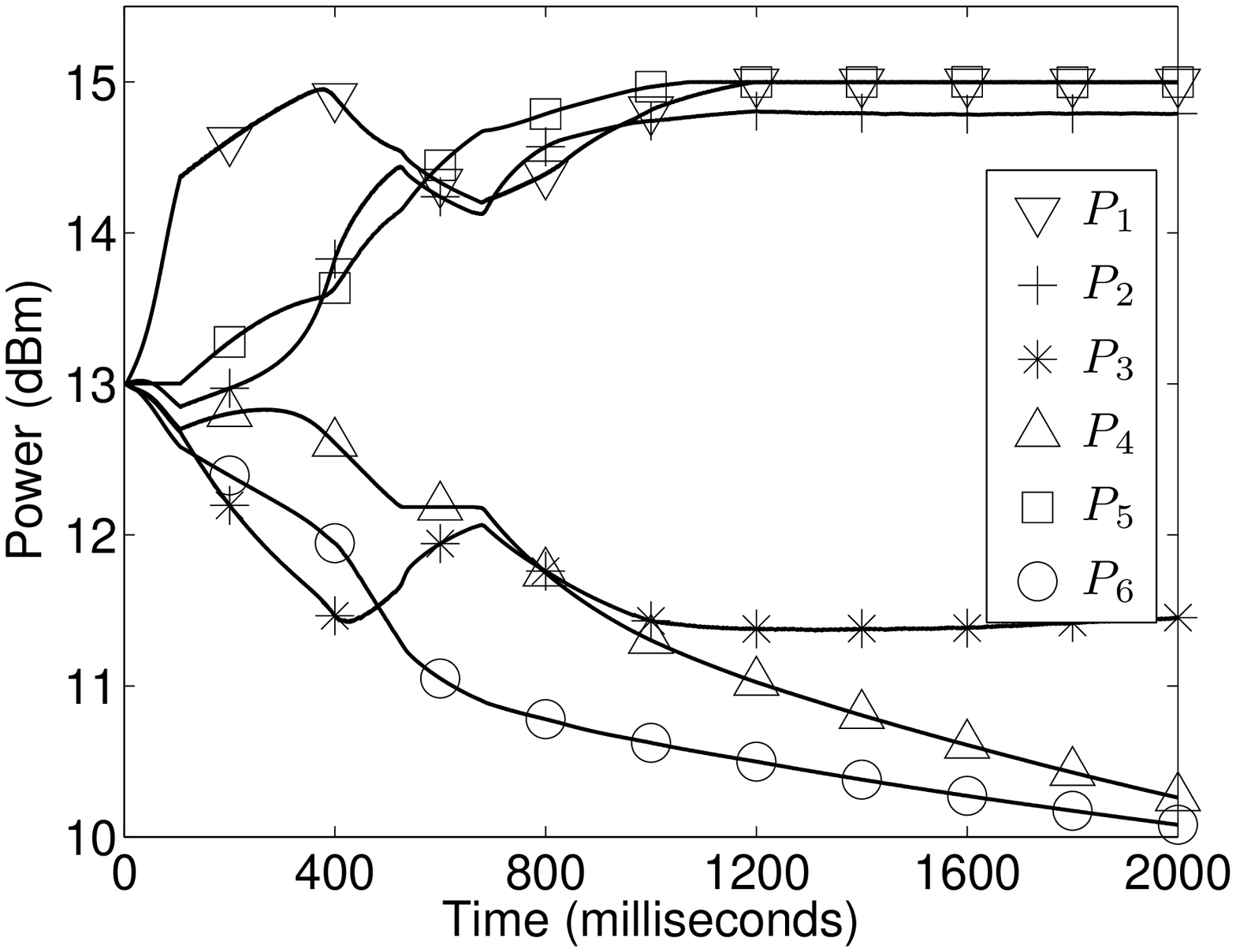}
	}
\caption{Centralized power control (a) Throughput, (b) power adjustment.}
\label{fig:pcc}
\end{center}
 \vspace{-0.3cm}
\end{figure}

Figure \ref{fig:pcc} shows the results of throughput and power adjustment of the centralized power control algorithm. In Figure \ref{fig:pcc-t}, we can see that without power control, the throughputs of the three destination nodes, node $4, 5, 6$ are all less than $0.5$pkt/ms, and the minimum of them is less than $0.25$pkt/ms. With dynamic power control, the throughput of all the nodes is improved. From around $t=400$ms, the throughputs begin to converge to the same value, $1$pkt/ms. Since each node, including the source node, has a transmission rate of $1$pkt/ms, the optimal multicast throughput is also $1$pkt/ms and the centralized algorithm achieves this optimum, thereby showing that this network coding aware power control regains the broadcast advantage. From Figure \ref{fig:pcc-p} we can see that the transmit power at each node is continuously adjusted by the power control algorithm. When the algorithm is started, i.e., $t=0$ms, powers of node $1, 2, 5$ are increased, on the other hand, other nodes' power are suppressed. When approaching throughput convergence, the algorithm sets the powers of nodes $1,2,5$ to about $15$dBm, and all the other nodes' powers to less than $12$dBm. 

We also compare the dynamic power control algorithm with a benchmark algorithm. In the benchmark algorithm, we formulate the max-min throughput problem for RNC based on the model of network flows \cite{dim-data}. For each destination $d \in \mc R$, there is a flow $f^d$ associated with it. To make the interference tractable with the network flow model, we consider specifically that this network uses TDMA for media access control and assume unicast communication. We further assume that each node $i$ uses the same power for transmissions on all its outgoing links. Let $t_{i,j}$ denote the time fraction that node $i$ allocates to link $(i,j)$ and $E$ denote the set of all the point-to-point links. 
Let $W_i$ be the throughput of node $i$, $N_i$ be the neighbor nodes of node $i$ which can transmit to or receive from node $i$. $\lambda_i=1$ is the rate that node $i$ is transmitting. Then an equivalent formulation of the max-min throughput problem described in equation (\ref{eq:max-min}) with respect to power control is given for the network flow model as follows:
\begin{align}
 \label{eq:max-min-flow}
& \underset{}{\text{maximize}}
&& W \\
\label{eq:opt-c1} & \text{subject to}
& & W < W_d, d\in \mc R\\ 
\label{eq:opt-c2} & & & W_d = \sum_{j \in N_1} f^d_{j,1} - \sum_{j \in N_1} f^d_{1,j}.\\
& & & \sum_{j \in N_i} f^d_{j,i} - \sum_{j \in N_i} f^d_{i,j} = 0, i \ne 1, d\\
\label{eq:opt-c4}& & & 0\le f^d_{i,j} \le t_{i,j}z_{i,j} \\
& & & \sum_{j\in N_i} t_{i,j} < 1\\
& & & t_{i,j} > 0\\
\label{eq:maxmin-zij}
& & & z_{i,j} = \lambda_i \cdot  \left(1-Q\left(\sqrt{\frac{P_{i}^{\mt{Tx}} \cdot h_{ji}}{\sum_{m\ne j,i}(P_{m}^{\mt{Tx}}\cdot h_{jm}/g_i) +\sigma^2}
}
\right)\right)^l \\ 
& \text{variables}
& & f_{i,j}^d, t_{i,j}, W_d, (i,j) \in E, d\in \mc R \notag\\ 
& && \mb P^{\text{Tx}}, W. \notag
\end{align}
In equation (\ref{eq:maxmin-zij}), $h_{ji}$ is the link gain and $g_i$ is the processing gain. Note that the above (benchmark) formulation is a straightforward extension of the linear programming formulation of network coding (see \cite{chr-fund}) since the objective function, i.e., equation (\ref{eq:max-min-flow}), and the constraints (\ref{eq:opt-c1})-(\ref{eq:opt-c4}) are equivalent to it. Feeding this formulation with the same topology and initial conditions as that are used in the dynamic power control simulation, a numerical nonlinear program solver yields the optimal value of $1$pkt/ms with Sequential Quadratic Programming (SQP) method \cite{jor-opt}. However, note that this algorithm can only be applied in a static fashion, i.e., it does not retain the dynamic characteristics of the power control in equation (\ref{eq:cpc}). Without taking into account the dynamic growth of ranks in RNC, any changes in the network would render a previously optimal allocation unsatisfactory and to make it work, a new allocation has to be computed. 

%

\subsubsection{Online algorithm}
\begin{figure}[t]
\begin{center}
	\subfigure[Effect of power control on throughput] {
		\label{fig:pcd-t}
		 \includegraphics[width=7.5cm]{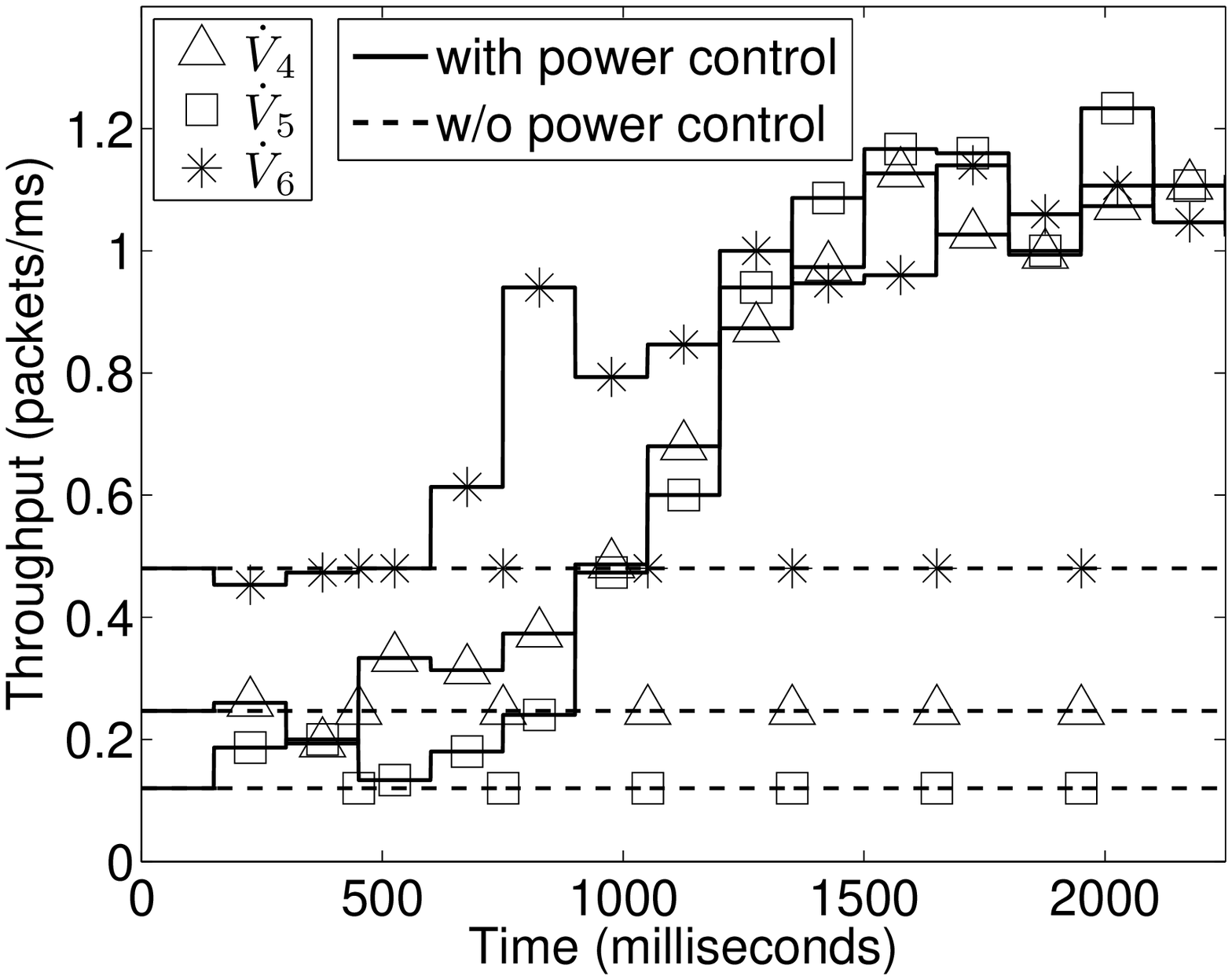}
	}
	\subfigure[Power adjustment] {
		\label{fig:pcd-p}
		 \includegraphics[width=7.5cm]{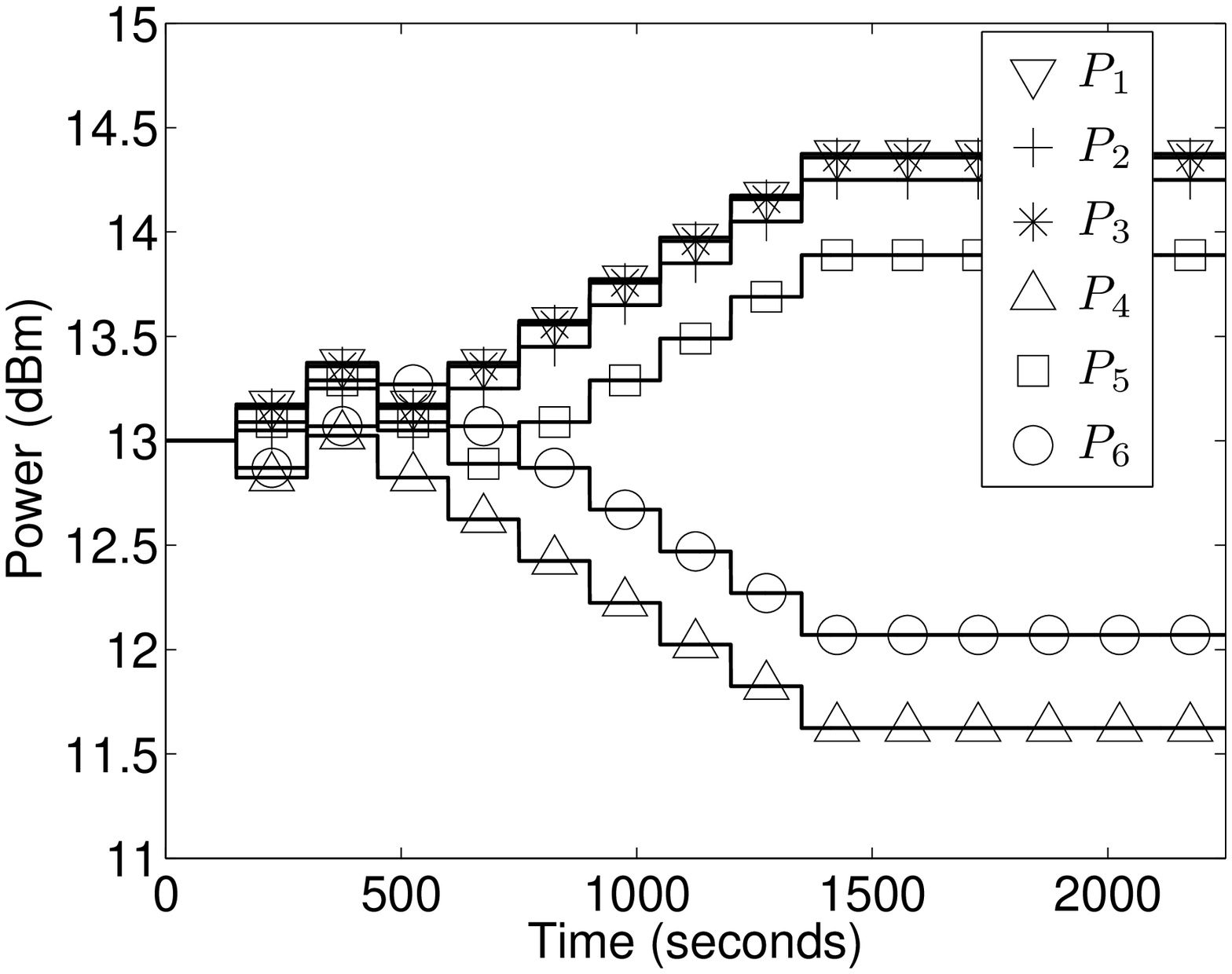}
	}

\caption{Online power control (a) Throughput, (b) power adjustment.}
\label{fig:pcd}
\end{center}
 \vspace{-0.3cm}
\end{figure}

In the simulation for online power control, the power is adjusted in every discrete time interval. Here, the interval is set to $150\mt{ms}$. The power levels of all the nodes are set to $13\mt{dBm}$ in the first interval, and a random update is applied in the second interval. After this initialization, in any interval $n$, we attempt to maintain a fixed step size $\gamma$ with the following rule:
\begin{equation}
P_i^{\mt{Tx}}(n) = \left \{ \begin{array}{ll}
					P_i^{\mt{Tx}}(n-1) + \gamma, & \text{if } a \cdot \frac{\partial T(n)}{ \partial P_i^{\mt{Tx}}}\ge\gamma  
				  \\
				P_i^{\mt{Tx}}(n-1) - \gamma, &\hfill \text{if } a \cdot \frac{\partial T(n)}{ \partial P_i^{\mt{Tx}}}\le-\gamma
				\\
				P_i^{\mt{Tx}}(n-1), & \text{otherwise.}
				\end{array} \right.	
\end{equation}
In this simulation, we set $\gamma$ as $0.2\mt{dBm}$. From Figure \ref{fig:pcd-t}, we can see in the first interval of this experiment, the destination nodes' largest throughput is about $0.5$pkt/ms, and the throughputs of node $4$ and $5$ are both less than $0.3$pkt/ms. As the algorithm progresses, the minimum throughput increases from about $0.1$pkt/ms to about $1$pkt/ms, and approaches convergence of about $1$pkt/ms around $t=1500$ms. Figure \ref{fig:pcd-p} shows how the transmit powers at each node are adjusted by the online power control algorithm.

\section{Dynamic CSMA Mean Backoff Delay Control in RNC} \label{sec:csma}
We now consider another resource allocation problem for RNC at the MAC layer, namely, CSMA mean backoff delay control. There exist a number of algorithms for maximizing CSMA throughput by link scheduling in a multihop network, as can be found in \cite{lib-csma,lib-csma2,jia-csma}. However, these algorithms do not consider or make use of the broadcast effect of wireless transmissions as RNC does. When CSMA is the underlying wireless media access control mechanism of a coded packet network, we perform dynamic CSMA backoff time control to optimize network coding performance, taking into account that each transmission is intrinsically a broadcast and neighbors will receive the transmitted packet. Although the general algorithm presented in section \ref{sec:algo} is sufficiently flexible such that it does not need to assume any particular model for $\lambda_i$, we now introduce a CSMA model for the purpose of illustration. We then apply the resource allocation algorithm in this setting to dynamically adjust the mean of backoff delay at each node to maximize the minimum throughput among the sink nodes. Note that the differential equation framework in equation (\ref{eq:dotv}) described earlier requires the knowledge of $z_{i, \mc K}$, the rate at which the packets from $i$ are successfully received by at least one node in $\mc K$. We will now derive $z_{i,\mc K}$ for this CSMA model.

\subsection{CSMA Model} \label{subsec:csmamodel}
The CSMA model considered here was first introduced in \cite{boo-csma} and is illustrative of a multihop network as is the case for RNC. We consider the wireless network $G=(\mc N, \mc E)$ and let $N_i$ denote the neighbors of node $i$, i.e., all the nodes that are within the range to be able to communicate with node $i$ and $N_i^* = N_i \cup \{i\}$. It should be noted that we assume here that the layer of network coding operation sits above the MAC layer on the protocol stack, and in this section, we refer to \textit{packet} as the coded packet with the MAC header attached. We describe next the assumptions for the CSMA model from \cite{boo-csma}:
\begin{itemize}
\item When a packet is scheduled to be transmitted at node $i$, the node senses the channel first. If the channel is idle, i.e., no ongoing transmissions from $N_i$, node $i$ transmits the packet immediately. We assume that the packet length is exponentially distributed with instantaneous transmission (zero propagation delay). We use $1/\mu_i$ to denote the mean of node $i$'s packet length.
\item After sensing the channel, if the channel is busy, node $i$ defers its transmission according to a random delay. The scheduling of packets, including the deferred ones, together with the newly scheduled after a successful transmission, is a Poisson process with rate $\alpha_i$. 
\item Any node from $N_i$, which are not currently receiving, can receive the packet from node $i$ without error immediately after the transmission. Subsequent transmissions heard by node $i$ while it is receiving will fail.
\end{itemize}

Let $x^i, i =0,1,...,M$ be a length-$N$ vector denoting the valid transmission status of the whole network where the $j$th component of $x^i$, $x^i_j$, is $1$ if node $j$ is transmitting, and $0$ otherwise. For instance, $[1\ 0\ 0\ 1\ 0\ 0]$ implies node $1$ and node $4$ are transmitting and other nodes are idle, and it is only valid when node $1$ and $4$ are not neighbors of each other. Let $G(x^i)$ denote the set of transmitting nodes in state $x^i$ and $H(x^i)$ denote the set of nodes that are not neighbors of any node in $G(x^i)$. The state transitions among all the states $x^i$ constitute a finite-state continuous Markov chain. Let $Q(x^i)$ be the stationary probability of state $x^i$, we have the following global balance equations:
\eq{
\begin{array}{ll}
& \sum_{j \in G(x^i)} Q(x^i) \mu_j + \sum_{j \in H(x^i)} Q(x^i) \alpha_j \\
&= \sum_{j \in G(x^i)} Q(x^i - \mb e_j) \alpha_j + \sum_{j \in H(x^i)} Q(x^i + \mb e_j) \mu_j,\ i=1, ..., M,
\end{array}
}
where $\mb e_j$ is a length-$N$ vector where the $j$th component is $1$ and $0$'s elsewhere. It can be verified that the following detailed balance equations hold:
\eq{
Q(x^i+\mb e_j) \mu_j = Q(x^i) \alpha_j, \ \ i=0,1,...,M, j \in H(x^i).
}

Let $x^0$ denote the state that no nodes are transmitting, and define $v_j = \alpha_j/\mu_j$. Then 
\eq{
Q(x^i) = Q(x^0)\prod_{j \in G(x^i)} v_j,\ \ i=1,...,M,
}
and 
\eq{
Q(x^0) = 1/(\sum_i \prod_{j \in G(x^i)} v_j).
}

Note that for a packet scheduled by node $i$ to be received by one of its neighbor nodes, $j$, the following requirements must be satisfied:
\begin{itemize}
\item $i$ must be idle. 
\item All the neighbors of $i$, including $j$, must be idle.
\item All the neighbors of $j$ must be idle. 
\end{itemize}
Based on the above results from \cite{boo-csma}, we can now proceed to derive the probability that a packet scheduled by node $i$ can be received by a neighbor node $j$ and let $P^{\prime}_{i,j}$ denote this probability. One should distinguish $P^{\prime}_{i,j}$ from $P_{i,j}$, since $P_{i,j}$ is the conditional probability that a packet \textit{transmitted} by node $i$ can be received by node $j$. Let $I(X)$ denote the event that all the nodes from the set $X$ are idle. Then
\eq{
P^{\prime}_{i,j} = \text{Prob}[I(N_i^* \cup N_j^*)] = \sum_{G(x^m) \subset \mc N \setminus (N_i^* \cup N_j^*)} Q(x^m).
}
Further, we continue to derive $P^{\prime}_{i,\mc K}$, the probability that a \textit{scheduled} packet from $i$ can be received by at least one node in $\mc K$. Let $\mc K^\prime = \mc K \cap N_i$, then
\eq{ \label{eq:csma-pik}
P^{\prime}_{i, \mc K} = P^{\prime}_{i, \mc K^\prime} =   \sum_{\exists \mc A \subset \mc K^{\prime}, \ \text{s.t.} \ G(x^m)\subset \mc N \setminus (N_i^* \cup N_{\mc A}^*) } Q(x^m),
}
where $N_{\mc A}^* = \cup_{i \in \mc A} N_i^* $.
Then, the rate at which node $i$'s packets are successfully received by at least one node from set $\mc K$ is given by:
\eq{ \label{eq:csma-zik}
z_{i,\mc K} = \alpha_i P^{\prime}_{i, \mc K}.
}

While the above developments assumed that packet length is exponentially distributed, it turns out that this assumption can be relaxed. For instance, it has been shown in \cite{sou-csma} that the results derived using the above model are more sensitive to the mean $\mu_i$, rather than the distribution itself. Therefore, the expressions in equations (\ref{eq:csma-pik}) and (\ref{eq:csma-zik}) are suitable for analyzing CSMA in a RNC network where transmitted data packets at the network layer are assumed to be of the same length.

\subsection{Centralized Gradient Algorithm for CSMA Mean Backoff Delay Control}
In a CSMA network, we consider the network resource $\mb r$ to be the rate of the Poisson scheduling process, $\bs \alpha$, i.e., $\mb r = \bs \alpha$. Since a Poisson process has exponentially distributed arrival times,  the backoff delay of node $i$ will be exponential with mean $1/\alpha_i$. Our goal is to adjust $\bs{\alpha}$ to maximize the minimum throughput. 
Thus, as long as we have the formulation of the CSMA model above, we know that the reception rate $z_{i,\mc K}$ is a function of $\bs{\alpha}$, i.e.,  
\eq{
z_{i,\mc K} = z_{i,\mc K} (\bs{\alpha}).
}
Let node $k$ be the node with the minimum throughput among all the destination nodes, i.e., $k = \arg \min_{j\in \mathcal{R}} \dot{V}_j$. Then based on the system of equations (\ref{eq:dotv}), we have:
\eq{
\begin{array}{ll}
\dot{V}_{k} &= \sum_{i\in N_k} z_{i,k}(1-q^{V_{k}-V_{\{i,k\}}}) \\
& = \sum_{i\in N_k} \alpha_i \cdot \left( \sum_{G(x^m) \subset \mc N \setminus (N_i^* \cup N_k^*)} Q(x^m) \right) \cdot
(1-q^{V_{k}-V_{\{i,k\}}})\\
& = \sum_{i\in N_k} \alpha_i \cdot \left( \sum_{G(x^m) \subset \mc N \setminus (N_i^* \cup N_k^*)} Q(x^0) \prod_{j\in G(x^m)} \alpha_j/\mu_j \right) \cdot
(1-q^{V_{k}-V_{\{i,k\}}}).
\end{array}
}
The versatility of the aforementioned dynamic resource allocation algorithm allows us to derive an algorithm customized for the CSMA backoff delay control problem. To this end, we follow equation (\ref{eq:gen-algo}) and get
\eq{ \label{eq:ccc}
\begin{array}{lll}
\dot \alpha_{i} & = a \cdot \left( \dot V_k(\bs \alpha + \Delta v\mb e_i) - \dot V_k(\bs \alpha) \right), \\ 
\end{array}
}
where $a$ is the control gain value, $\Delta v$ is the step size, and $\mb e_i$ is a vector whose $i$th component is $1$ with $0$ elsewhere.

\subsection{Online Gradient Algorithm for CSMA Mean Backoff Delay Control}
We can also derive an online version of the resource allocation algorithm for CSMA mean backoff delay control based on the centralized gradient algorithm. As in the case of dynamic power control, the online algorithm for CSMA backoff control at each node makes the estimation of the gradient without any knowledge of the underlying PHY/MAC layer specifics at other nodes, and iteratively adjusts the value of the mean backoff delay $\bs \alpha$:
\begin{equation} \label{disc-c}
\bs{\alpha}(n) = \bs{\alpha}(n-1) + a \cdot \nabla_{\bs{\alpha}}T(n-1),
\end{equation}
where $\bs{\alpha}(n)$ is $\bs \alpha$ at the $n$th time interval.
Note that the gradient to be approximated is $\nabla_{\bs{\alpha}}T{(n)}$. We can utilize a similar derivation as for the case of power control which results in: 
\eq{ \label{grad-disc-c}
\nabla_{\bs{\alpha}}T(n) = 
\nabla_{\dot{\mathbf{V}}}T(n) 
J_{\mathbf{z}^{\prime}}\dot{\mathbf{V}} (n) 
J_{\bs{\alpha}}\mathbf{z}^{\prime}(n),
}
where $J_{\bs{\alpha}}\mathbf{z}^{\prime}$ is the Jacobian of $\mathbf{z}^{\prime}$ with respect to $\bs{\alpha}$.

To reduce the computational overhead, we can still follow Broyden's method \cite{broyden} to estimate $J_{\bs{\alpha}}\mathbf{z}^{\prime}(n)$
\eq{ \label{broyden-c}
J_{\bs{\alpha}}\mathbf{z}^{\prime} (n) = J_{\bs{\alpha}}\mathbf{z}^{\prime}(n-1) + 
\frac{\Delta\mathbf{z}^\prime(n) - J_{\bs{\alpha}}\mathbf{z}^{\prime}(n-1) \Delta \bs{\alpha}(n)}{\| \Delta \bs{\alpha}(n)\|^2}
\Delta \bs{\alpha}^\top(n).
}



\subsection{Numerical Results}
We perform two kinds of evaluations: (i) a differential equation solver based evaluation for centralized CSMA mean backoff delay control and (ii) an event-driven simulation for the online algorithm. We use the $6$-node network shown in Figure \ref{fig:concept_topo} and consider that each node has a set of neighbor nodes, as summarized in Table \ref{tab:node}. The neighbors are determined based on a thresholding distance between nodes.
\begin{table}[h]
\centering
\begin{tabular}{|c|c|}
\hline
Node & Neighbors\\
\hline
$1$ & $\{3, 5\}$ \\
\hline
$2$ & $\{5, 6\}$ \\
\hline
$3$ & $\{1, 4, 5\}$ \\
\hline
$4$ & $\{3, 5, 6\}$ \\
\hline
$5$ & $\{1, 2, 3, 4, 6\}$ \\
\hline
$6$ & $\{2, 4, 5\}$ \\
\hline
\end{tabular}
\caption{Nodes and their neighbors.} \label{tab:node}
\end{table}
Recall that $\alpha_i$ is the scheduling rate of node $i$. We measure this rate in packets per millisecond (pkt/ms). 
In both evaluations, we introduce a scalar parameter $\beta_i$ taking the value of $\log \alpha_i$. We call $\beta_i$ the transmission aggressiveness of node $i$ and set $3\le \beta_i \le 6.5, i=1,2,...,N$. Initially, $\beta_i$ of every node $i$ is set to $4$ and therefore the average scheduling rate $\alpha_i$ is about $0.06$pkt/ms. We assume the packets are of fixed length and set $1/\mu_i$ of any node $i$ to be $1/1000$ms. 
\subsubsection{Centralized algorithm}
\begin{figure}[t]
\begin{center}
	\subfigure[Effect of contention window size control on throughput] {
		\label{fig:ccc-t}
		 \includegraphics[width=7.5cm]{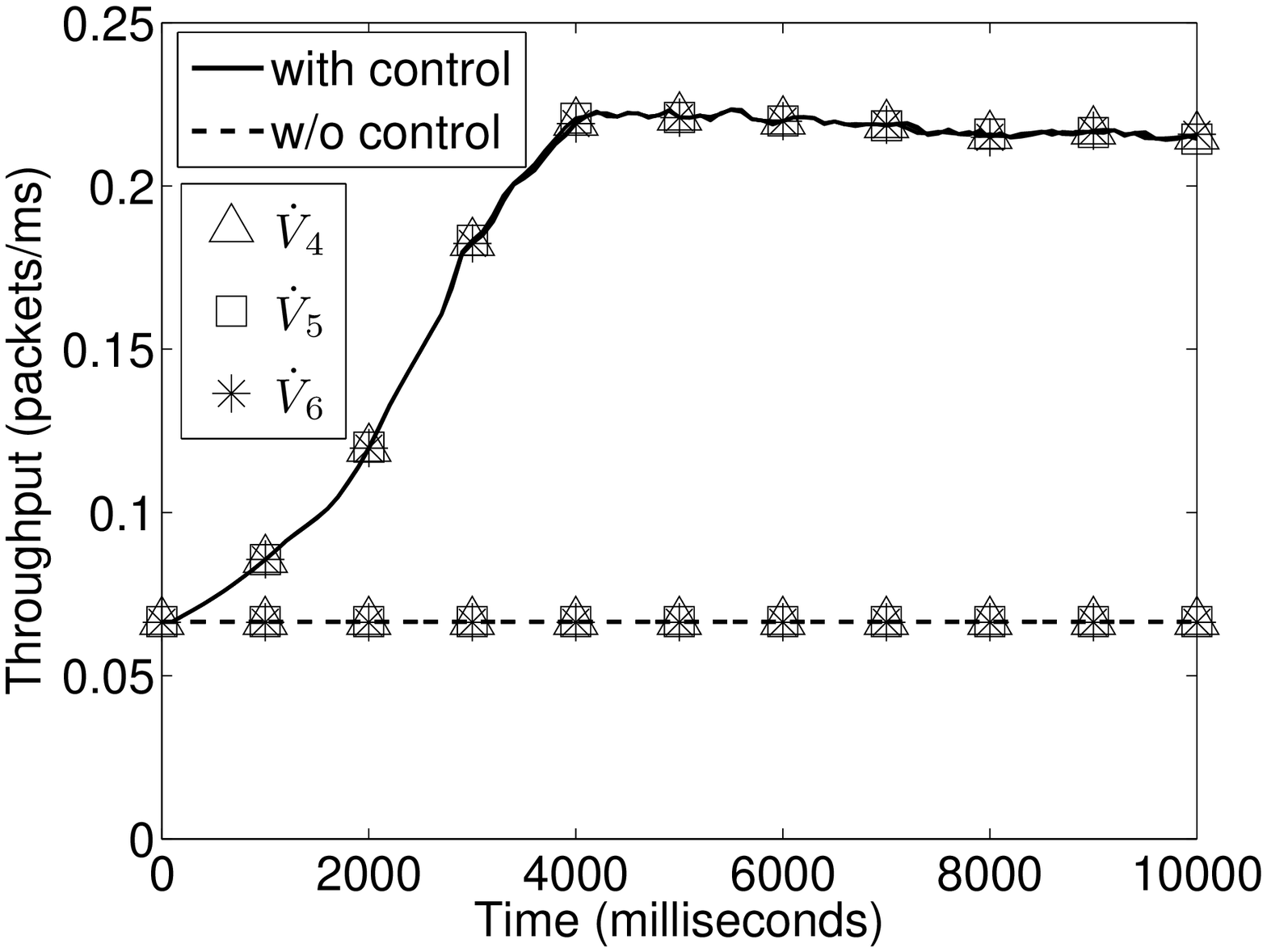}
	}
	\subfigure[Contention window size adjustment] {
		\label{fig:ccc-b}
		 \includegraphics[width=7.5cm]{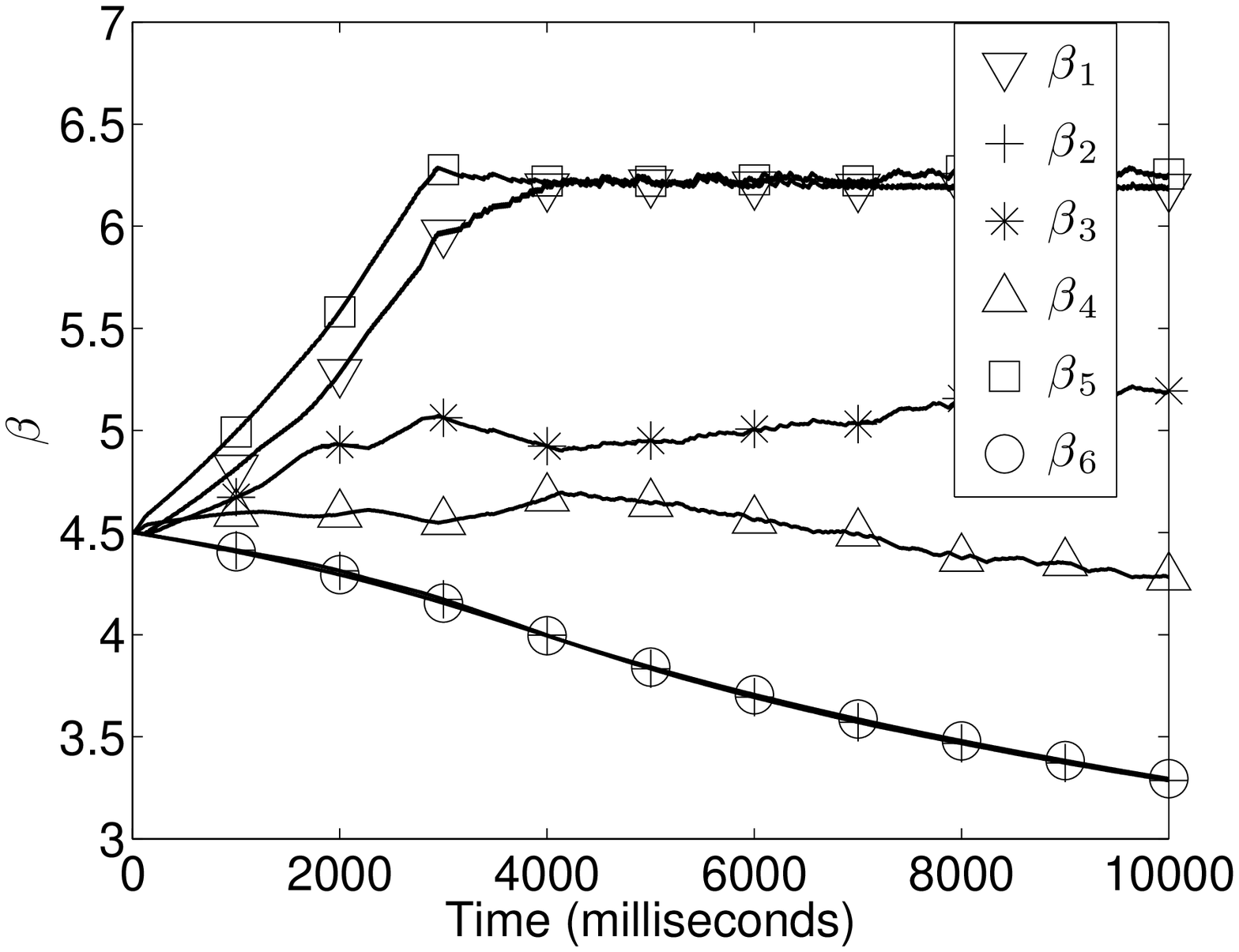}
	}

\caption{Centralized contention window size control (a) Throughput, (b) contention window size adjustment.}
\label{fig:ccc}
\end{center}
 \vspace{-0.3cm}
\end{figure}

The results of DE solver based evaluation for centralized CSMA mean backoff delay control is shown in Figure \ref{fig:ccc}. The dashed lines in Figure \ref{fig:ccc-t} indicate that without dynamic backoff control, the throughput of nodes $4, 5, 6$ remain at about $0.06$pkt/ms. The resource allocation algorithm increases all the three destination nodes' performance and their throughputs converge to about $0.22$pkt/ms, making it a gain of greater than $200\%$ from the initial throughputs. Figure \ref{fig:ccc-b} shows how the transmission aggressiveness $\beta$ is continuously adjusted at each node. 

\subsubsection{Online algorithm}
\begin{figure}[t]
\begin{center}
	\subfigure[Effect of contention window size control on throughput] {
		\label{fig:ccd-t}
		 \includegraphics[width=7.5cm]{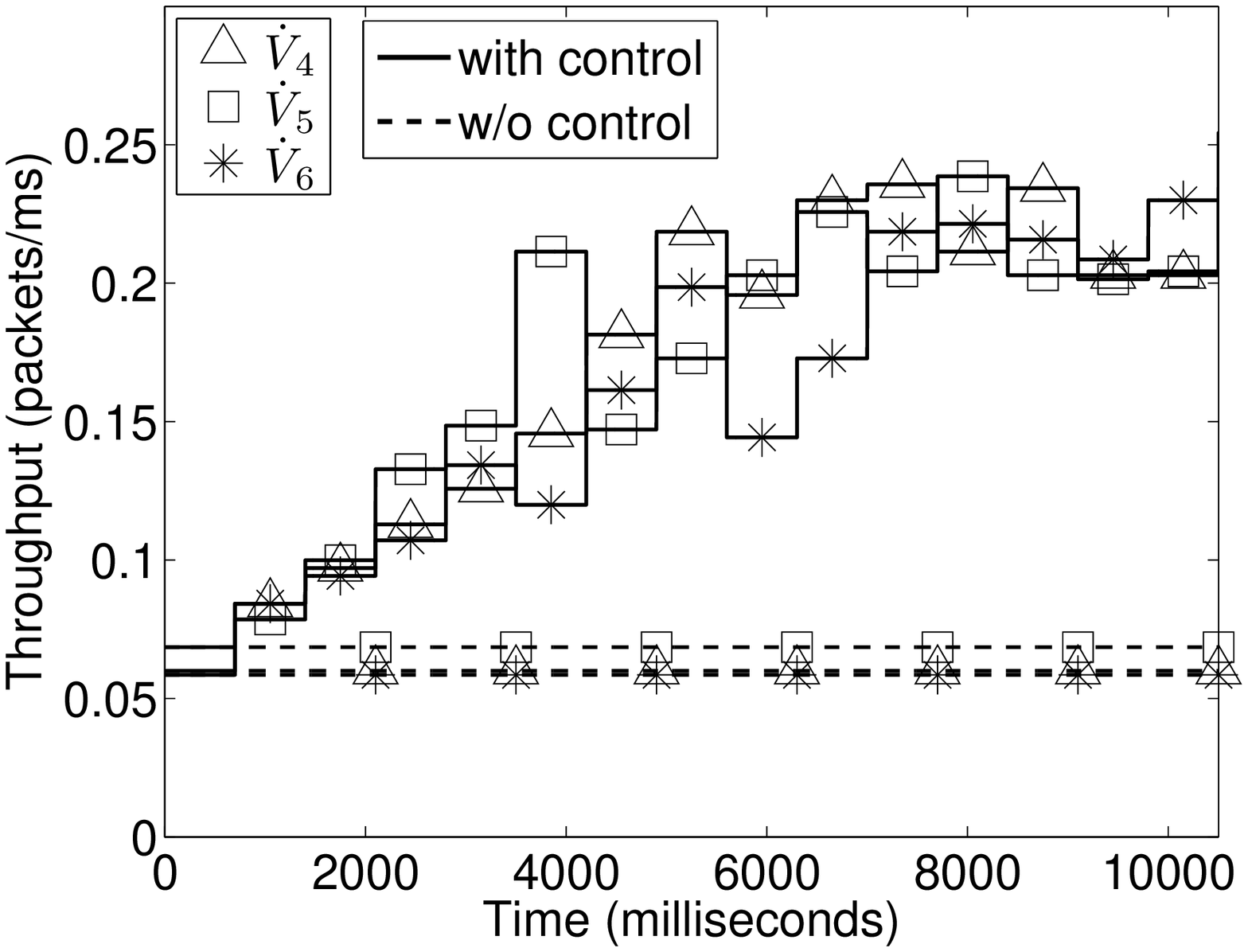}
	}
	\subfigure[Contention window size adjustment] {
		\label{fig:ccd-b}
		 \includegraphics[width=7.5cm]{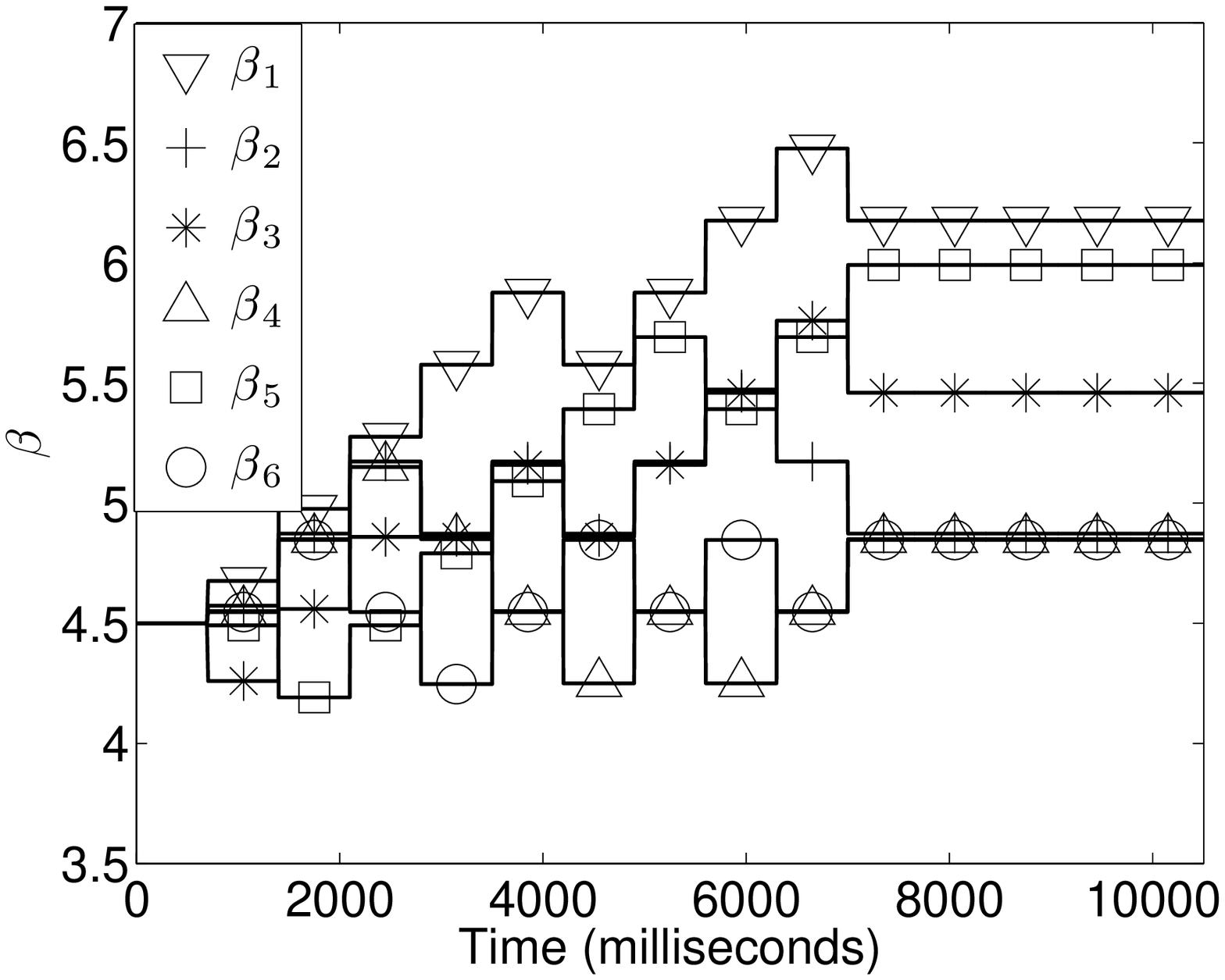}
	}

\caption{Online contention window size control (a) Throughput, (b) contention window size adjustment.}
\label{fig:ccd}
\end{center}
 \vspace{-0.3cm}
\end{figure}

From Figure \ref{fig:ccd-t}, we can see that all the three destinations' original throughputs are less than $0.075$pkt/ms at $t=0$ms. Starting from $t=1$ms, the online resource allocation algorithm gradually improves the throughputs of all the destination nodes. Around $t=7000$ms, the throughputs begin to converge around $0.20$pkt/ms. Figure \ref{fig:ccd-b} shows how the transmission aggressiveness is adjusted by the online algorithm.

\section{Conclusion}
We investigated integrated resource allocation for wireless networks which employ random network coding as the transport scheme. We used a differential equation based framework that models RNC throughput, thereby enabling the analysis of RNC performance in terms of PHY and MAC layer parameters. Using this framework, we designed dynamic power control and CSMA mean backoff delay control algorithms to improve the performance of RNC. Specifically, we used gradient based resource allocation algorithms and evaluated both centralized and online versions of them, via the use of differential equation solvers and event driven simulations. Our results revealed that such network coding aware resource allocation significantly improves the throughput of destination nodes in RNC. We also observed that such integrated power control can regain the broadcast advantage. Beyond the use cases of power control and CSMA backoff control, the framework and approach presented in this paper can be generally applied to a variety of resource allocation problems for RNC.





\bibliographystyle{IEEEtran}
\bibliography{kai-ref}

\begin{thebibliography}{10}
\providecommand{\url}[1]{#1}
\csname url@samestyle\endcsname
\providecommand{\newblock}{\relax}
\providecommand{\bibinfo}[2]{#2}
\providecommand{\BIBentrySTDinterwordspacing}{\spaceskip=0pt\relax}
\providecommand{\BIBentryALTinterwordstretchfactor}{4}
\providecommand{\BIBentryALTinterwordspacing}{\spaceskip=\fontdimen2\font plus
\BIBentryALTinterwordstretchfactor\fontdimen3\font minus
  \fontdimen4\font\relax}
\providecommand{\BIBforeignlanguage}[2]{{%
\expandafter\ifx\csname l@#1\endcsname\relax
\typeout{** WARNING: IEEEtran.bst: No hyphenation pattern has been}%
\typeout{** loaded for the language `#1'. Using the pattern for}%
\typeout{** the default language instead.}%
\else
\language=\csname l@#1\endcsname
\fi
#2}}
\providecommand{\BIBdecl}{\relax}
\BIBdecl

\bibitem{tan-pc}
J.~Tang, G.~Xue, C.~Chandler, and W.~Zhang, ``Link scheduling with power
  control for throughput enhancement in multihop wireless networks,'' \emph{in
  IEEE Transactions on Vehicular Technology}, vol.~55, no.~3, pp. 733 -- 742,
  May 2012.

\bibitem{geo-realloc}
L.~Georgiadis, M.~Neely, and L.~Tassiulas, ``Resource allocation and
  cross-layer control in wireless networks,'' \emph{Foundations and
  Trends\textregistered\ in Networking}, vol.~1, no.~1, 2006.

\bibitem{xia-realloc-suv}
X.~Lin, N.~Shroff, and R.~Srikant, ``A tutorial on cross-layer optimization in
  wireless networks,'' \emph{Selected Areas in Communications, IEEE Journal
  on}, vol.~24, no.~8, pp. 1452 --1463, aug. 2006.

\bibitem{mung-realloc-suv}
M.~Chiang, S.~Low, A.~Calderbank, and J.~Doyle, ``Layering as optimization
  decomposition: A mathematical theory of network architectures,''
  \emph{Proceedings of the IEEE}, vol.~95, no.~1, pp. 255 --312, jan. 2007.

\bibitem{netflow}
R.~Ahlswede, N.~Cai, S.-Y. Li, and R.~Yeung, ``Network information flow,''
  \emph{IEEE Transactions on Information Theory}, vol.~46, no.~4, pp. 1204
  --1216, July 2000.

\bibitem{han-sec}
K.~Han, T.~Ho, R.~Koetter, M.~Medard, and F.~Zhao, ``On network coding for
  security,'' \emph{IEEE Military Communications Conference (MILCOM)}, pp. 1 --
  6, Oct. 2007.

\bibitem{dima-dist-stor}
A.~Dimakis, P.~Godfrey, Y.~Wu, M.~Wainwright, and K.~Ramchandran, ``Network
  coding for distributed storage systems,'' \emph{Information Theory, IEEE
  Transactions on}, vol.~56, no.~9, pp. 4539 --4551, sept. 2010.

\bibitem{gkan-cont-dist}
C.~Gkantsidis and P.~Rodriguez, ``Network coding for large scale content
  distribution,'' in \emph{INFOCOM 2005. 24th Annual Joint Conference of the
  IEEE Computer and Communications Societies. Proceedings IEEE}, vol.~4, march
  2005, pp. 2235 -- 2245 vol. 4.

\bibitem{yal-cros}
Y.~Sagduyu and A.~Ephremides, ``Cross-layer optimization of mac and network
  coding in wireless queueing tandem networks,'' \emph{Information Theory, IEEE
  Transactions on}, vol.~54, no.~2, pp. 554 --571, feb. 2008.

\bibitem{dan-rand}
D.~Traskov, D.~S. Lun, R.~Koetter, and M.~Medard, ``Network coding in wireless
  networks with random access,'' in \emph{Information Theory, 2007. ISIT 2007.
  IEEE International Symposium on}, june 2007, pp. 2726 --2730.

\bibitem{zha-reallocisit}
D.~Zhang, K.~Su, and N.~B. Mandayam, ``Network coding aware resource allocation
  to improve throughput,'' \emph{IEEE International Symposium on Information
  Theory (ISIT)}, 2012.

\bibitem{bene}
T.~Ho, R.~Koetter, M.~Medard, D.~Karger, and M.~Effros, ``The benefits of
  coding over routing in a randomized setting,'' \emph{in IEEE International
  Symposium on Information Theory}, p. 442, Jul. 2003.

\bibitem{dedi-it}
D.~Zhang and N.~B. Mandayam, ``Analyzing random network coding with
  differential equations and differential inclusions,'' \emph{IEEE Transactions
  on Information Theory}, vol.~57, no.~12, pp. 7932--7949, Dec. 2011.

\bibitem{dedi-isit}
D.~Zhang, N.~Mandayam, and S.~Parekh, ``\uppercase{DEDI}: A framework for
  analyzing rank evolution of random network coding in a wireless network,'' in
  \emph{Information Theory Proceedings (ISIT), 2010 IEEE International
  Symposium on}, june 2010, pp. 1883 --1887.

\bibitem{oncoding}
D.~S. Lun, M.~Medard, R.~Koetter, and M.~Effros, ``On coding for reliable
  communication over packet networks,'' \emph{in 42nd Annual Allerton
  Conference on Communication, Control, and Computing}, Sept. 2004.

\bibitem{rus-opt}
A.~Ruszczynski, \emph{Nonlinear Optimization}.\hskip 1em plus 0.5em minus
  0.4em\relax Princeton University Press, 2006.

\bibitem{mun-pc}
M.~Chiang, P.~Hande, T.~Lan, and C.~Tan, ``Power control in wireless cellular
  networks,'' \emph{Foundations and Trends\textregistered\ in Networking},
  vol.~2, no.~4, April 2008.

\bibitem{fos-dpc}
G.~Foschini and Z.~Miljanic, ``A simple distributed autonomous power control
  algorithm and its convergence,'' \emph{Vehicular Technology, IEEE
  Transactions on}, vol.~42, no.~4, pp. 641 --646, nov 1993.

\bibitem{yates-pc}
R.~Yates, ``A framework for uplink power control in cellular radio systems,''
  \emph{Selected Areas in Communications, IEEE Journal on}, vol.~13, no.~7, pp.
  1341 --1347, sep 1995.

\bibitem{zand-pc}
J.~Zander, ``Distributed cochannel interference control in cellular radio
  systems,'' \emph{Vehicular Technology, IEEE Transactions on}, vol.~41, no.~3,
  pp. 305 --311, aug 1992.

\bibitem{chiang-uti-pc}
M.~Chiang and J.~Bell, ``Balancing supply and demand of bandwidth in wireless
  cellular networks: utility maximization over powers and rates,'' in
  \emph{INFOCOM 2004. Twenty-third AnnualJoint Conference of the IEEE Computer
  and Communications Societies}, vol.~4, march 2004, pp. 2800 -- 2811 vol.4.

\bibitem{hand-utit-pc}
P.~Hande, S.~Rangan, M.~Chiang, and X.~Wu, ``Distributed uplink power control
  for optimal sir assignment in cellular data networks,'' \emph{Networking,
  IEEE/ACM Transactions on}, vol.~16, no.~6, pp. 1420 --1433, dec. 2008.

\bibitem{mand-pc}
C.~Saraydar, N.~Mandayam, and D.~Goodman, ``Efficient power control via pricing
  in wireless data networks,'' \emph{Communications, IEEE Transactions on},
  vol.~50, no.~2, pp. 291 --303, feb 2002.

\bibitem{ciss}
K.~Su, D.~Zhang, and N.~B. Mandayam, ``Network coding aware power control in
  wireless netoworks,'' \emph{in 46th Annual Conference on Information Sciences
  and Systems (CISS)}, 2012.

\bibitem{broyden}
C.~G. Broyden, ``A class of methods for solving nonlinear simultaneous
  equations,'' \emph{Mathematics of Computation (American Mathematical
  Society)}, Oct. 1965.

\bibitem{itu}
``Propagation data and prediction methods for the planning of indoor radio
  communication systems and the radio local area networks in the frequency
  range 900 mhz to 100 ghz,'' \emph{ITU-R Recommendations}, 2001.

\bibitem{dim-data}
D.~P. Bertsekas, \emph{Network Optimization: Continuous And Discrete
  Models}.\hskip 1em plus 0.5em minus 0.4em\relax Athena Scientific., 1998.

\bibitem{chr-fund}
C.~Fragouli and E.~Soljanin, ``Network coding fundamentals,'' \emph{Foundations
  and Trends\textregistered\ in Networking}, vol.~2, no.~1, 2007.

\bibitem{jor-opt}
J.~Nocedal and S.~J. Wright, \emph{Numerical Optimization}.\hskip 1em plus
  0.5em minus 0.4em\relax Springer., 2006.

\bibitem{lib-csma}
L.~Jiang and J.~Walrand, ``A distributed csma algorithm for throughput and
  utility maximization in wireless networks,'' \emph{Networking, IEEE/ACM
  Transactions on}, vol.~18, no.~3, pp. 960 --972, june 2010.

\bibitem{lib-csma2}
------, ``Approaching throughput-optimality in distributed csma scheduling
  algorithms with collisions,'' \emph{Networking, IEEE/ACM Transactions on},
  vol.~19, no.~3, pp. 816 --829, june 2011.

\bibitem{jia-csma}
J.~Ni and R.~Srikant, ``Distributed csma/ca algorithms for achieving maximum
  throughput in wireless networks,'' in \emph{Information Theory and
  Applications Workshop, 2009}, feb. 2009, p. 250.

\bibitem{boo-csma}
R.~Boorstyn, A.~Kershenbaum, B.~Maglaris, and V.~Sahin, ``Throughput analysis
  in multihop csma packet radio networks,'' \emph{Communications, IEEE
  Transactions on}, vol.~35, no.~3, pp. 267 -- 274, mar 1987.

\bibitem{sou-csma}
S.~C. Liew, C.~H. Kai, H.~C. Leung, and P.~Wong, ``Back-of-the-envelope
  computation of throughput distributions in csma wireless networks,''
  \emph{Mobile Computing, IEEE Transactions on}, vol.~9, no.~9, pp. 1319
  --1331, sept. 2010.

\end{thebibliography}

\end{document}